# From Twitter to traffic predictor: Next-day morning traffic prediction using social media data

Weiran Yao [a], Sean Qian [a,b,*]

[a] Department of Civil and Environmental Engineering, Carnegie Mellon University, Pittsburgh, PA 15213, United States
[b] Heinz College, Carnegie Mellon University, Pittsburgh, PA 15213, United States

A B S T R A C T

The effectiveness of traditional traffic prediction methods, such as autoregressive or spatio-temporal models, is often extremely limited when forecasting traffic dynamics in early morning. The reason is that traffic can break down drastically during the early morning commute, and the time and duration of this break-down vary substantially from day to day. Early morning traffic forecast is crucial to inform morning-commute traffic management, but they are generally challenging to predict in advance, particularly by midnight (called 'next-day morning traffic prediction' thereafter). In this paper, we propose to mine Twitter messages as a probing method to understand the impacts of people's work and rest patterns in the evening/midnight of the previous day to the next-day morning traffic. The model is tested on freeway networks in Pittsburgh as experiments. The resulting relationship is surprisingly simple and powerful. We find that, in general, the earlier people rest as indicated from Tweets, the more congested roads will be in the next morning. The occurrence of big events in the evening before, represented by higher or lower tweet sentiment than normal, often implies lower travel demand in the next morning than normal days. Besides, people's tweeting activities in the night before and early morning (by 5am) are statistically associated with congestion in morning peak hours. We make use of such relationships to build a predictive framework which forecasts morning commute congestion using people's tweeting profiles extracted by 5am. In most cases, the tweet information collected by the midnight before is sufficient to make good prediction for next-day morning traffic. The Pittsburgh study supports that this framework can precisely predict morning congestion, particularly for some road segments upstream of roadway bottlenecks with large day-to-day congestion variation, while its prediction performance being no worse than baseline methods on other roads. Through experiments, we demonstrate our approach considerably outperforms those existing methods without Twitter message features, and it can learn meaningful representation of demand from tweeting profiles that offer managerial insights. The proposed social media empowered framework can be a promising tool for real-time traffic management and potentially extended for traffic prediction at other times of day.

## 1. Introduction

Existing traffic prediction methods are often of limited use to early morning commuters. According to American Community Survey

* Corresponding author at: Department of Civil and Environmental Engineering, Carnegie Mellon University, Pittsburgh, PA 15213, United States.
*E-mail address:* seanqian@cmu.edu (S. Qian).

(2011–2015) by U.S. Census Bureau (2015), 13% of the population nationwide were reported to leave home for work before 6am to avoid the worst commute times, and 4.4% were even out the door by 5am. Congestion on roadways is known be extremely sensitive to travel demand. On each day, the departure time and volumes of those early morning commuters could cause traffic break down at various times of day, which has ripple effects on the following morning commuters. From day to day, morning commute would result in completely different congestion patterns on some road segments (as will be seen in the examples later). Therefore, as the goal of this study, a model capable of forecasting traffic during the entire morning commute (e.g. 5-10am) for those road segments (particularly those primary ones) before early morning (e.g. by 5am) or as late as midnight, would be mostly desired. Only with the ahead-of-curve prediction on when, where roads will start to congest and how long it would last, traffic managers can proactively engage information dissemination and various operational strategies.

Unfortunately, characteristics of free-flow traffic prevalent in transportation networks before the early morning are uninformative and do not necessarily represent congestion patterns during the morning rush hours. As illustrated in Fig. 1(a), autoregression-based method (AR) with a lag of 5 or 10 min is often unable to pick up the traffic break-down using only the time-series traffic data up to 5am, except for a small fraction of days when heavy delay appeared already earlier than 5am. Also, historical information does not seem to help due to the high variance of congestion patterns from day to day as shown in Fig. 1(b). Consequently, traditional traffic prediction models, which primarily rely on correlations between future and real-time traffic speed data, could potentially fail for such a next-day morning traffic prediction setting. This limitation by the lack of explanatory data, however, can be best tackled by employing data sources implying broader impacts on morning rush hour traffic. Our overarching idea is to use general data implying population general activities during the night before and early morning by 5am to predict traffic patterns of the entire morning commute period, called 'next-day morning traffic prediction' thereafter.

Following traffic flow theory, traffic congestion occurs when travel demand exceeds available link supply (Sheffi, 1985). While supply-side descriptors, such as weather or traffic events (e.g., road construction, bridge precaution, etc.) can be easily collected, daily travel demand data is almost inaccessible on a daily basis. Historical travel surveys cannot account for the day-to-day variance of travel demand. Most crowdsourcing techniques (e.g., wearable sensors, mobile phones, etc.) which are capable of tracking and tracing individual activities before early morning (Harrison et al., 2020) are often insufficient for sensing the entire metropolitan area, nor do they offer predictive insights on congestion patterns. Questions arise: (1) is there a pervasive computing tool for collecting population's night and early morning general activities in regional networks that would strongly imply travel characteristics in the morning commute? and (2) can we mine and translate the collected data into morning traffic predictors by 5am on each day?

Recently, the rise of social media provides new opportunities to understand the relationship between people's daily activities and urban systems. User-generated contents, together with account profiles, posting time, and locations can be readily available in real-time with Internet access. Tweets posted in the night and early morning may imply people's activity patterns, such as sleep-wake statuses or travel plans on the next day. Some demonstrative example tweets are listed in Table 1.

This paper presents `tweet2traffic`, a new class of social computing models for learning meaningful representation of morning travel demand directly from Twitter messages and using it to improve next-day morning traffic prediction in transportation networks. The proposed models integrate state-of-the-art language representation (e.g. neural language model, sentiment analysis) and geo-processing techniques to map each tweet information into a high dimension feature space, and extracts daily tweeting profiles through aggregation by space and time. Specifically, we argue that tweets capture three types of useful information for explaining next-day morning traffic, which includes people's sleep-wake status, local events, and (planned) traffic incidents showcased in Table 1. Considering the sparsity of geocoded tweets in the night before and early morning, we propose a novel social media data augmentation method which first filters influential users (i.e., residents) from noisy users, such as tourists, infers their approximate home locations, and augments the dataset with user timelines such as non-geocoded tweets, retweets, and favorites by assuming residents stay at home during the night before. A sentiment analysis model with a neural language model as a backend is constructed to detect abnormal local events, and incident records are extracted from traffic-related tweets.

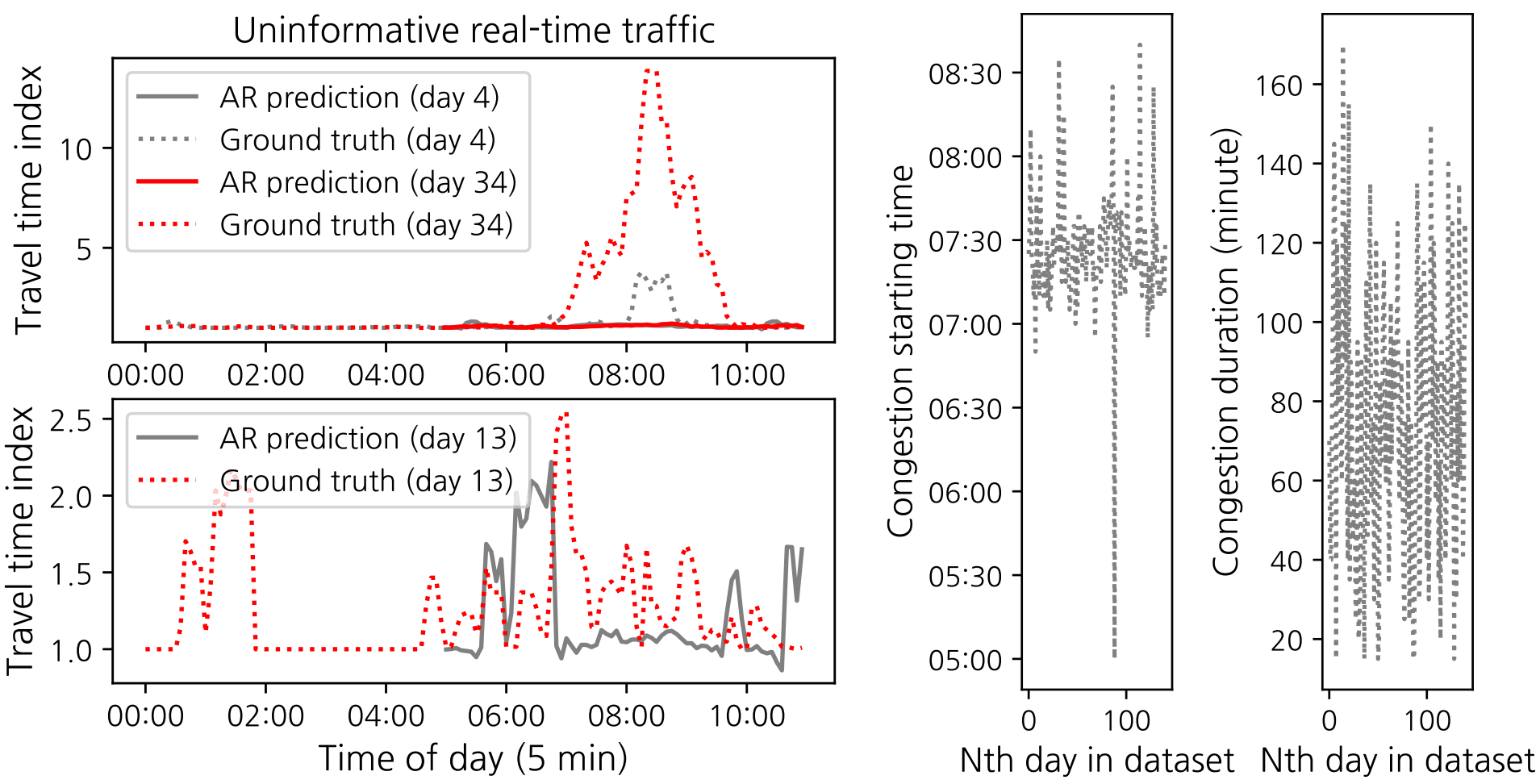

(a) Autoregression-based (AR) method. (b) History-based method.

**Fig. 1.** Actual and predicted travel time of a representative highway road segment in Pittsburgh (ID:104-04531): time series data prior to 5am is generally uninformative.

**Table 1**
Tweet examples capturing three types of information useful for next-day morning traffic.

| | |
|---|---|
| Sleep-wake status (sleep) | "Good night [Sleeping face] I'm off to bed, but before I go, remember I love you." |
| Sleep-wake status (wake) | "Got up at 4am to take some friends to the airport. Def bout to take a nap" |
| Local event | "Not a good night for the @Pirates but it was great to spend some time near PNC Park with @TheBuccosFan @mryannagy @ANTOINETTE180 @padavies3." |
| (Planned) traffic incident | "Short-term lane restrictions on Route 22 in Robinson Township, will begin on Friday, July 31 through August 21 between the 7AM to 5PM. Both east and westbound directions between the I-576 and the Bavington exits will be restricted." |

Using freeway networks in the Pittsburgh metropolitan area as a case study, the proposed models encode the supply-side variables (traffic incidents, weather, etc.) and demand-side variables (tweet profiles, temporal variables, etc.) through a clustered learning structure which makes use of ordered spatio-temporal congestion patterns of segments along a road. Through experiments, we demonstrate that our approach outperforms existing methods without social media features for morning traffic prediction in terms of congestion starting time, duration, and planning time index prediction errors, and can learn meaningful representations of travel demand from tweeting profiles. Generally, we discover that the earlier people go to sleep, the more congested the road will be in the next morning. Tweeting activities in the early morning by 5am are positively associated with congestion in the morning peak hours. Also, we find that the occurrence of big events in the evening before, represented by higher or lower tweet sentiment than normal, often results in lower travel demand in the next morning than normal weekdays. We illustrate through ablation studies that social media data, traffic events, and weather conditions are important data sources for next-day morning traffic prediction and their roles vary by space and time. We find that tweet related features can largely improve traffic prediction for road segments near the edge of spillback segments of a traffic bottleneck, where travel demand plays a crucial role in determining congestion. Besides, the implications of social media are more significant when special events such as sports games occur during the night, which possibly reduces or puts off morning travel demand. For traffic incidents and weather conditions, their impacts are obvious on non-recurrent days with roadworks or adverse weather conditions. Traffic incidents downstream to road segments generally shift congestion starting time while upstream incidents reduce congestion duration. Adverse weather conditions such as snow, rain, and floods often make roads more congested by bringing forward congestion starting time and increasing congestion duration and scales. Finally, our approach proves insensitive to the increase of forecasting horizons even before midnight and outperforms benchmarks using the same length of traffic or tweet data without the clustered structure.

### *1.1. Related work*

*Transportation research and social media.* Transport operators have recently incorporated social media into their situation awareness strategies (Cottrill et al., 2017; Rashidi et al., 2017). Promising transportation applications on social media can be categorized into traffic related incident detection (Zhang et al., 2018; Suma et al., 2017; Gu et al., 2016; D'Andrea et al., 2015; Schulz et al., 2013), social event detection (Khare et al., 2019), trip purpose (Cui et al., 2018; Gkiotsalitis and Stathopoulos, 2015) and travel demand modeling (Markou et al., 2019; Berlingerio et al., 2017; Hu and Jin, 2017), human mobility exploration (Zhang et al., 2017; Zhao and Zhang, 2017; Hasan and Ukkusuri, 2014), travel information retrieval (Kuflik et al., 2017) and so on. A detailed review of recent work can be found in (Zhang and He, 2019). Another branch of research used geotagged social media as traveler reviews and explored relations between public opinions, represented by tweet check-in types or counts and content, with infrastructure design (Huang et al., 2017; Ali et al., 2017). On the contrary, little has been done using the information from social media to directly inform real-time traffic prediction. Steiger et al. (2016) explored the spatiotemporal relationship between mobility patterns with traffic-related tweets using self-organizing maps. The strong correlations between spatiotemporal tweet clusters with proximity to special events, traffic incidents, and hazard reports show that social media can serve as a proxy indicator of collective mobility events and predict short-term traffic during unplanned events. Zhang et al. (2016) studied the correlation of traffic-related tweets with traffic surges within certain spatial and temporal ranges around posting coordinates and timestamps. Ni et al. (2016) used the counts of tweet users and event-related tweets to predict passenger flow at a subway station under event conditions. Lin et al. (2015) extracted adverse weather features from tweets and used it for traffic speed prediction. However, prior work either examined near-term or lagged effects of social media on traffic or focused on specific events, improving long-term prediction of day-to-day traffic with social media was not explored. The closest work to this is the work of (He et al., 2013), which used tweet posting activity and bag-of-words semantic features to enhance the prediction of traffic volumes averaged over the next 1,2,12 and 24 h. Although one of their forecasting horizons (12 h) is similar to our study, the time resolution of predicted traffic is insufficient for morning commuters. The spatial information of tweets and user profiles were not used. When, where, and how can social media patterns and semantics during the night and early morning improve next-day morning traffic prediction were also unexamined.

*Traffic prediction.* While the literature concerning specifically on morning traffic prediction is not vast, recent studies show a trend of increased forecasting horizon to better inform traffic management. Previous work focused on short-term traffic prediction (e.g., 5 min to 15 min ahead) and their methods do not generalize well to forecasting horizons beyond 30 min. Autoregressive models (Williams and Hoel, 2003; Smith and Demetsky, 1997), Kalman filter (Guo et al., 2014) and non-parametrics (Sun et al., 2003) are some common modeling choices. A detailed review of short-term traffic prediction can be found in (Oh et al., 2015). Recently, spatiotemporal relations of network traffic have been explored in literature to achieve accurate and ahead-of-curve real-time prediction. Recent work either incorporates spatiotemporal priors explicitly through constraints (Min and Wynter, 2011), or develops learnable deep learning architecture (Ma et al., 2017; Cui et al., 2020; Polson and Sokolov, 2017; Yao and Qian, 2020; Yang et al., 2019) which encodes inductive bias from network flow physics. Comprehensive reviews of recent work in spatiotemporal traffic prediction can be found in (Wang et al., 2020; Xie et al., 2020; Ermagun and Levinson, 2018). However, spatiotemporal methods don't work for next-day morning traffic prediction since free-flow traffic is prevalent in transportation network before early morning and no meaningful spatiotemporal signal can thus be extracted to predict traffic of the whole morning period. Another branch of research increased forecasting horizon with data from interdependent civil systems. Lagged dependency between building occupancy changes and road traffic nearby was examined in (Zheng et al., 2016). HVAC data were used to approximate building occupancy and as features to infer longer-term traffic states. Zhang and Qian (2018) found household electricity usage impacted highway congestion during morning peak hours. Weather and urban events (Yang and Qian, 2019) have also been employed as explanatory variables for future network traffic. However, no existing work that we are aware of mines Twitter messages in evening/midnight to predict next-day morning traffic prediction on daily basis.

*Social media analytics.* Social media departs from other GPS-enabled crowdsourcing techniques in two ways: (1) social media data only include geographical positions at sparse points in time and the pathway between these points is unknown, and (2) user-generated contents and account profiles come with each data point. Therefore, state-of-the-art social media analytics focuses on mining spatiotemporal patterns from check-in coordinates and opinions embedded in contents. Clustering methods such as K-means (Kanungo et al., 2002), Gaussian Mixture Model or GMM (Banfield and Raftery, 1993), Density-Based Spatial Clustering of Applications with Noise or DBSCAN (Ester et al., 1996) and Mean-Shift (Comaniciu and Meer, 2002) are the most popular choices for identifying spatial clusters from tweet posting coordinates and exploring human mobility patterns (França et al., 2016; Wu et al., 2014; Hasan et al., 2013). Recent literature (Lin and Cromley, 2018; Huang et al., 2016) successfully inferred user home locations from just a few geocoded tweets without supervision. Our study is built upon DBSCAN and features (e.g., check-in rates) in these works for locating user home locations. For information retrieval from social media contents, keyword-based approach is mostly used, where a domain-specific dictionary is built beforehand and can be expanded later through the Apriori algorithm (Agarwal et al., 1994). Sentiment analysis derives people's opinions and sentiment from a piece of writing, classifying it as is positive, negative or neutral (Liu, 2012). Commonly-used methods include lexicon-based (Bakshi et al., 2016; Hu and Liu, 2004) and machine learning approaches (Agarwal et al., 2011). Topic modeling can embed a tweet into a vector of topic features where each topic is represented by relevant word distributions. Popular topic modeling approaches are Latent Dirichlet Allocation or LDA (Blei et al., 2003) and Nonnegative Matrix Factorization or NMF (Shahnaz et al., 2006). Recently, deep neural language models (LM) via self-supervision have produced several pre-trained models for general-purpose language understanding, which include BERT (Devlin et al., 2019), OpenAI GPT-3 (Brown et al., 2020) and so on. Neural language models project a document of variable length into a high-dimensional vector while keeping semantics and local structure of the document. Our work is built upon BERT to extract sentiment from tweets for detecting local events.

### *1.2. The contribution of this study*

Our paper is differentiated from prior work in four ways:

1. We propose a novel machine learning model to use social media data to directly improve the morning traffic prediction on daily basis. This problem is fundamentally different from prior work that can be classified in the following three types: usage of social media for traffic incident detection or travel demand modeling, the examination of near-term effects of social media on traffic, or studying specific events;
2. Three types of traffic-related information including people's sleep-wake status, local events, and (planned) traffic incidents are defined and extracted from tweets. A novel social media data augmentation method that tackles the sparsity of geocoded tweets at night and early morning is proposed to supplement the dataset by geotagging non-geocoded tweets during this period with inferred user home locations. Those pieces of information can improve the prediction accuracy tremendously;
3. Meaningful travel demand representation from tweeting profiles is extracted through learning this next-day morning traffic prediction. This is a data driven approach different from most existing works constructing complex and hypothetical steps to approximate travel demand;
4. Our study illustrates the different roles of social media, traffic incidents, and weather conditions in traffic prediction by space and time. When, where and how the incorporation of multi-source data would improve traffic prediction offer implementation guidelines to transportation operators and "interprets" the black box of machine learning.

## 2. Dataset

The main objective of this paper is to use tweets posted during the night before and early morning to improve the prediction of morning traffic congestion. This section describes the data sources used in the paper: (1) traffic datasets, which are comprised of probe-sourced traffic speed data from INRIX Traffic[1] and Road Condition Report System (RCRS) incident data[2] maintained by the Pennsylvania Department of Transportation; (2) weather datasets from Weather Underground[3]; and (3) Twitter streaming datasets from the free Twitter Streaming API[4]. Data within the City of Pittsburgh, Pennsylvania from Jan 23, 2014 to Dec 31, 2014 were collected.

### 2.1. INRIX speed dataset

Historical traffic speed data were obtained from INRIX which cover major US highways in Pittsburgh metropolitan. The traffic speed data were georeferenced by Traffic Message Channel (TMC) coding. The raw datasets spanned 4,254 TMCs in the City of Pittsburgh. Since this paper focuses on morning traffic congestion, only 53 major highway TMC segments, where congestion (defined as the travel time index $\geqslant 2$) is observed in the morning periods (from 5am to 11am) at least once in the analysis period, are considered. The simplified morning transportation network with 53 TMCs along I-279 Southbound (I-279 S), PA-28 Southbound (PA-28 S), I-376 Westbound (I-376 W), and I-376 Eastbound (I-376 E) are shown in Fig. 2. All selected TMCs, not surprisingly, are heading from suburban areas to downtown Pittsburgh. The road speed data were reported every 5 min. Each record includes TMC code, timestamp, observed speed (mph), average speed (mph), reference speed (mph) and two parameters for the confidence of the speed, namely confidence score and confidence value.

### 2.2. PennDOT RCRS incident dataset

RCRS data feeds provide real-time information for traffic incidents, roadwork, winter road conditions, and other events that cover all state-owned roads. RCRS incident records include the incident location, road closure and open time stamps, as well as the category of the incident. In 2014, RCRS reported 2,696 traffic incidents in Pittsburgh. Fig. 2 shows the incidents in our simplified network by location and category. It is found that most incidents are roadwork, which can be assumed to be planned ahead of time.

### 2.3. Weather underground dataset

Weather underground reported hourly weather measurements. Each entry contains temperature, pressure, dew point, humidity, wind speed, precipitation, pavement condition, and visibility, etc. Weather data collected in Pittsburgh International Airport were used, and the location of weather station is shown in Fig. 2.

### 2.4. Twitter dataset

We construct our Twitter streaming dataset by collecting all geocoded and non-geocoded tweets posted within the bounding box (−80.20, 40.29; −79.80, 40.62) visualized in Fig. 2. 1,782,636 tweets from Jan 23, 2014 to Dec 31, 2014 were collected. This is done by first inquiring Twitter Streaming APIs using this bounding box to retrieve 1,349,179 geocoded tweets from 43,670 users, and then furthermore collecting user timeline tweets posted by those 43,670 users through Twint[5], a Twitter web scraper. Those users were used to retrieve additional non-geocoded tweets and favorites, adding up to all 1,782,636 raw tweets including both geocoded and non-geocoded. Of those raw tweets, 672,527 (37.72%) tweets posted by those 43,670 users inquired from Twitter Streaming APIs are tagged with accurate locations. The Twitter data include date/time, text, user ID, language, latitude and longitude (if available), user profile location, etc. Also, 2014 US Census Tract Cartographic Boundary Shapefiles are used to join geocoded tweets with the geographical information.

## 3. Descriptive analysis: feasibility of using tweets for next-day traffic prediction

We first conduct a study of the correlation between the geocoded Twitter data from late night to 5am and the morning traffic data for the roads in our simplified transportation network. This can be seen as a proof of concept to show the feasibility of our study. To examine the correlation in a straightforward way, we characterize the common patterns of morning traffic congestion and geocoded tweeting activities using cluster analysis. The relationship between traffic and geocoded tweeting patterns are visualized and tested by chi-squared statistics. The design of descriptive analysis was inspired by (Li et al., 2016; Li et al., 2015).

### 3.1. Clustering of morning traffic

We follow the steps illustrated in Fig. 3 to identify typical morning traffic congestion patterns of a road. First, we transform the raw speed data into Travel Time Index ($TTI_{it}^{d}$) to measure congestion on TMCs. The Travel Time Index is defined as the ratio of travel time in the peak period to the travel time at free-flow conditions (FHWA, 2019). Thus, $TTI_{it}^{d}$ on a road segment $i$ at time index $t$ on day $d$ can be computed by Eq. (1), as the ratio of the free-flow (reference) traffic speed $v_i^{ref}$ to the observed speed $v_{it}^{d}$ on segment $i$. To determine the reference (free-flow) speed $v_i^{ref}$ of a TMC $i$, the 85 percentile of observed speed on that segment for all time periods (Eq. (2)) is used, which is the recommended approach for computing reference speed from probe-based speed data (Jha et al., 2018).

[1] INRIX traffic: https://inrix.com/products/ai-traffic/.
[2] RCRS incident: https://www.penndot.gov/Doing-Business/OnlineServices/Pages/Developer-Resources.aspx.
[3] Weather underground: https://www.wunderground.com/.
[4] Twitter streaming API: https://developer.twitter.com/en/docs.
[5] Twint: https://github.com/twintproject/twint.

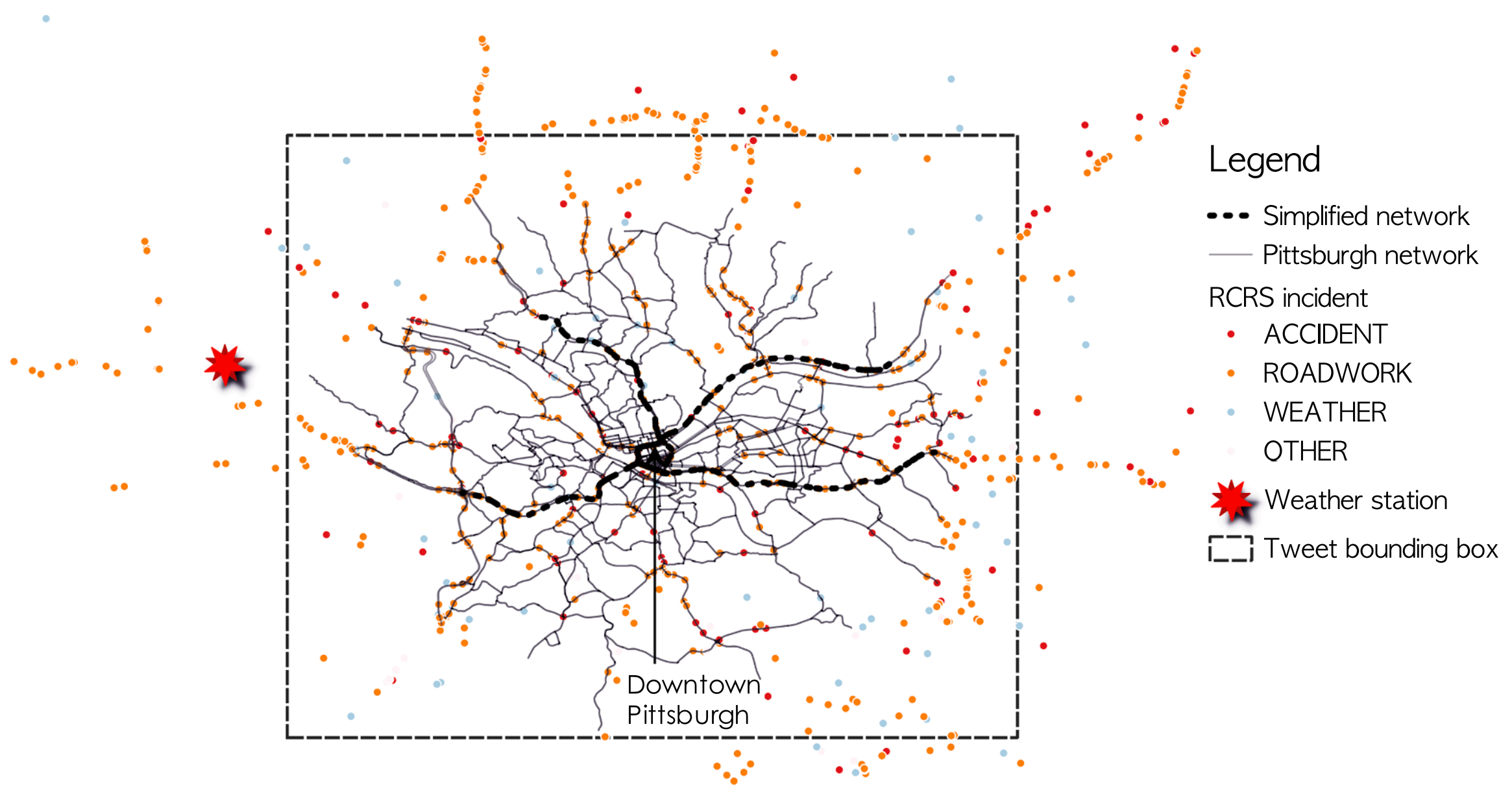

**Fig. 2.** Datasets used in this paper.

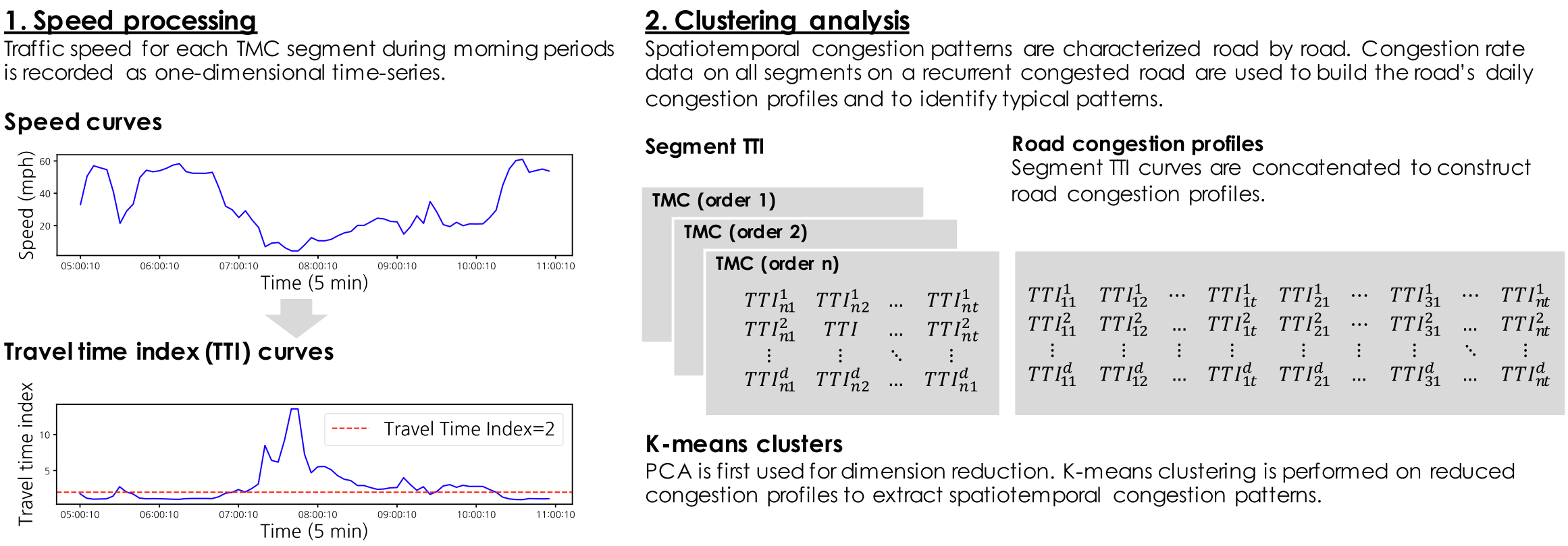

**Fig. 3.** Speed processing steps for clustering morning congestion.

$$TTI_{it}^{d} = v_i^{ref} / v_{it}^{d} \tag{1}$$

$$v_i^{ref} = \mathbb{P}_{0.85}(v_{it}^{d}) \tag{2}$$

Then, we generate daily traffic congestion profiles of a road, by concatenation of $TTI_{it}^{d}$ from 5am to 11am, for all TMCs on that road. A daily profile for road becomes a vector of $\mathbb{R}^{N\times T}$, where $N$ denotes the number of TMCs on that road and $T = 72$ is the number of sampling points of morning periods (in 5-min time intervals). In this study, we adopt principle component analysis (PCA) to address the sparseness and multicollinearity of high-dimension spatiotemporal congestion profiles. To remove noise while keeping most information, the first $P$ components that capture 90% of the variance are kept. Finally, clustering analysis, using K-means on the reduced $P$-dimension matrix, is conducted to identify typical daily morning congestion patterns for each road in the network. Optimal cluster size $K$ is selected by the elbow method.

The morning road traffic congestion profiles show ordered spatio-temporal patterns, as presented in Fig. 4 where the traffic clusters are ordered from top to bottom by average $TTI$ of the cluster centroid. Because of spill-back effects, severe congestion with greater $TTI$ usually has more spatial and temporal impacts. Surprisingly, we also find that congestion profiles with greater *TTI* are often coupled with earlier congestion starting time on the road. This might suggest that how early commuters hit the road have some impacts on morning congestion patterns. We intend to use social media data the night before to directly learn about the early-morning demand behavior so as to better interpret/predict the morning congestion.

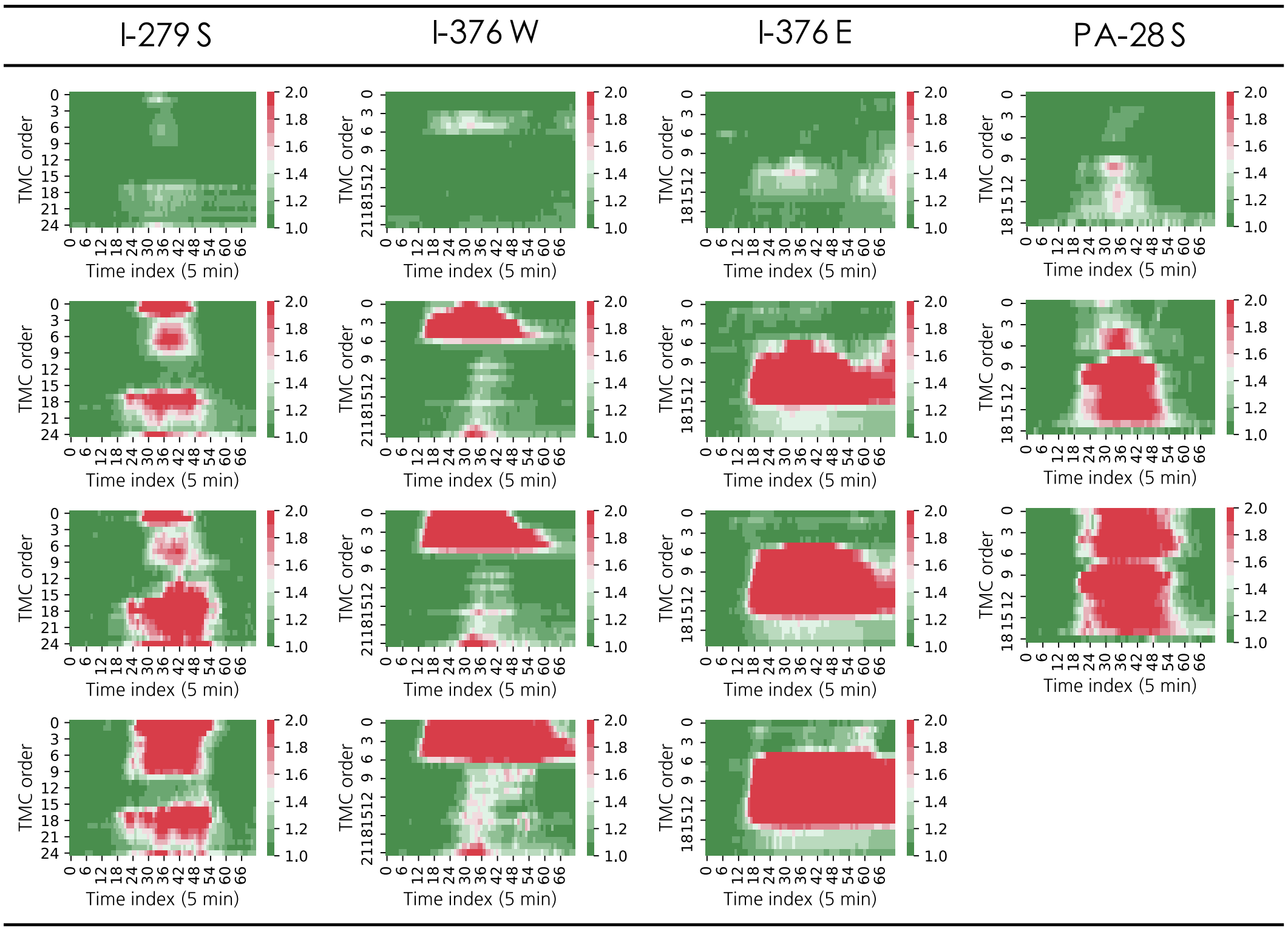

**Fig. 4.** Traffic cluster centers ordered by average travel time index (TTI).

### 3.2. Characterization of tweeting activities during the night before and early morning

One intuition for using tweets from late night to early morning to predict morning traffic is that tweeting activities can imply daily sleep-wake related activities in the regional network and might be used to probe how early and how many commuters hit the road, which consequently, as illustrated in Fig. 4, impacts morning congestion patterns. An example would be if some users tweet late at 2am, they cannot or don't have the plan to get up early and add to morning traffic.

This section characterizes such sleep-wake patterns captured by geocoded tweets with cluster analysis. For simplicity, we remove spatial information of the tweets for the time being, i.e., all tweets within the bounding box are seen as posted from the same Pittsburgh region. To build tweeting profiles of each day, tweets posted from 6 pm to 5am are first selected, and their counts are grouped by day and (half) hour index. To remove noise, a moving average smoother of a two-hour window is applied. Then, we normalize the tweeting activity curve by its total number, so that each daily profile sums up to one. A daily tweeting profile becomes a vector of $\mathbb{R}^{19}$. Finally, similar K-means clustering setups in Section 3.1 are used to derive typical sleep-wake patterns captured by tweets.

Four representative tweeting activity clusters are identified from the data. As shown in Fig. 5, while most tweets were posted between 8:30 and 10:30 pm, when the peak is reached and how tweeting activities declined after the peak vary substantially among clusters, and from day to day. To illustrate this, cluster 1 shows intense tweeting activities in midnight. Cluster 2 implies steady decrease in activities till early morning, which differs from a dramatic decline to low tweeting activities immediately after 10:30 pm in clusters 1, 3 and 4. Clusters 3 and 4 show intense tweeting activities in the evening and late night, but cluster 4 reaches a peak earlier before 9:00 pm, and its activities are more concentrated around the peak.

### 3.3. Correlation analysis

We visualize and analyze the correlation by conditioning traffic congestion cluster distribution on tweeting activity patterns with 300 days of data. Fig. 6 implies a surprisingly simple relationship for four roads in different spatial locations in Pittsburgh, that the earlier tweeting activity profile achieves the peak during the night (tweeting peak: clusters $4 < 3 < 2 < 1$), the more likely severe congestion (traffic clusters: $4 > 3 > 2 > 1$ or $3 > 2 > 1$) occurs in the next-day morning. The intuition is that average sleep time of the population can be inferred from tweeting activity peaks (since apparently a person can't tweet when they are asleep). As a result, because the earlier a person goes to sleep, the more likely they get up in the morning early and need to make the commuting trip, both of which can increase morning traffic congestion. Hence, tweeting activities are likely to be transformed into a probing measurement of morning travel demand. It is also observed that the more concentrated the tweeting activity is (cluster 4), the more likely congestion is to occur. This also suggests the feasibility of using Twitter messages as a representation of morning travel demand.

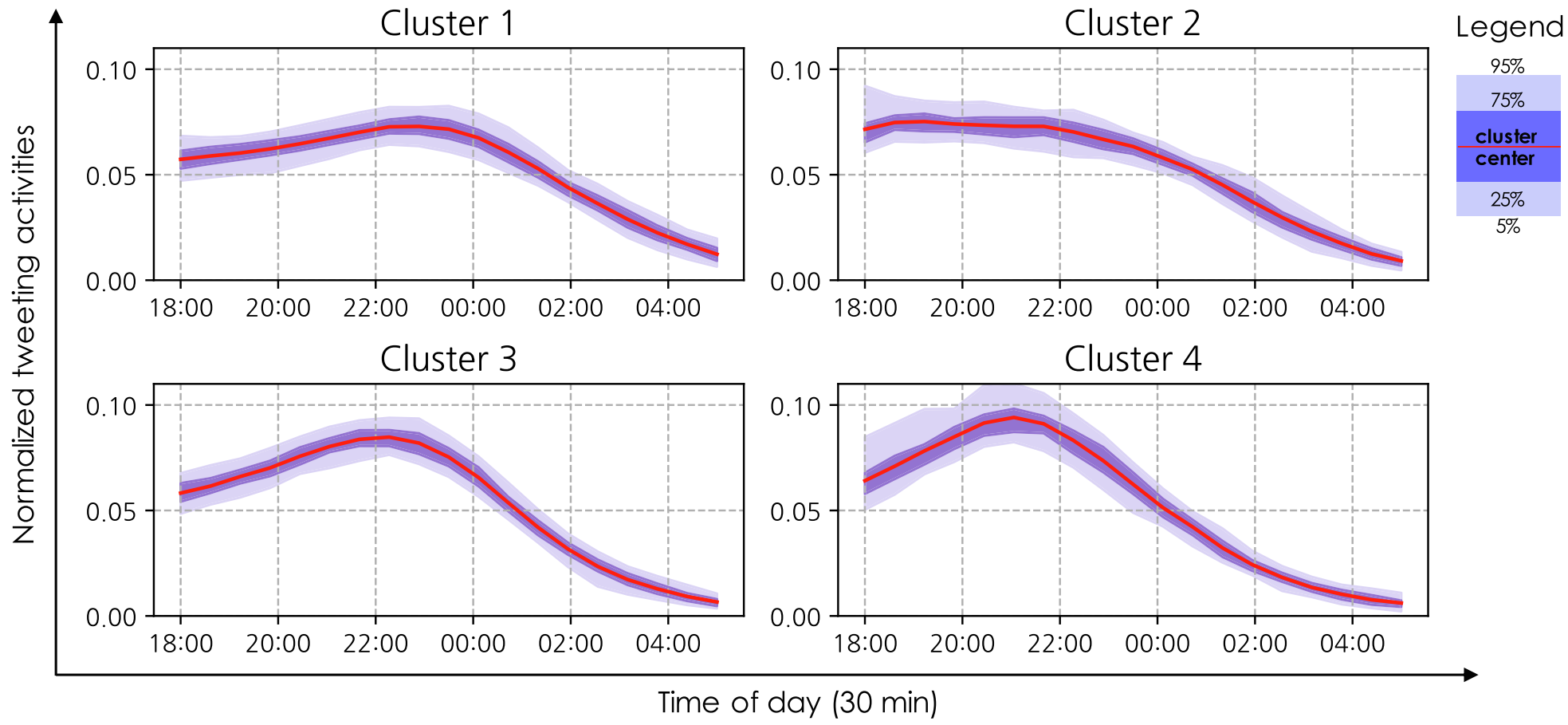

**Fig. 5.** Typical late night and early morning tweeting profiles.

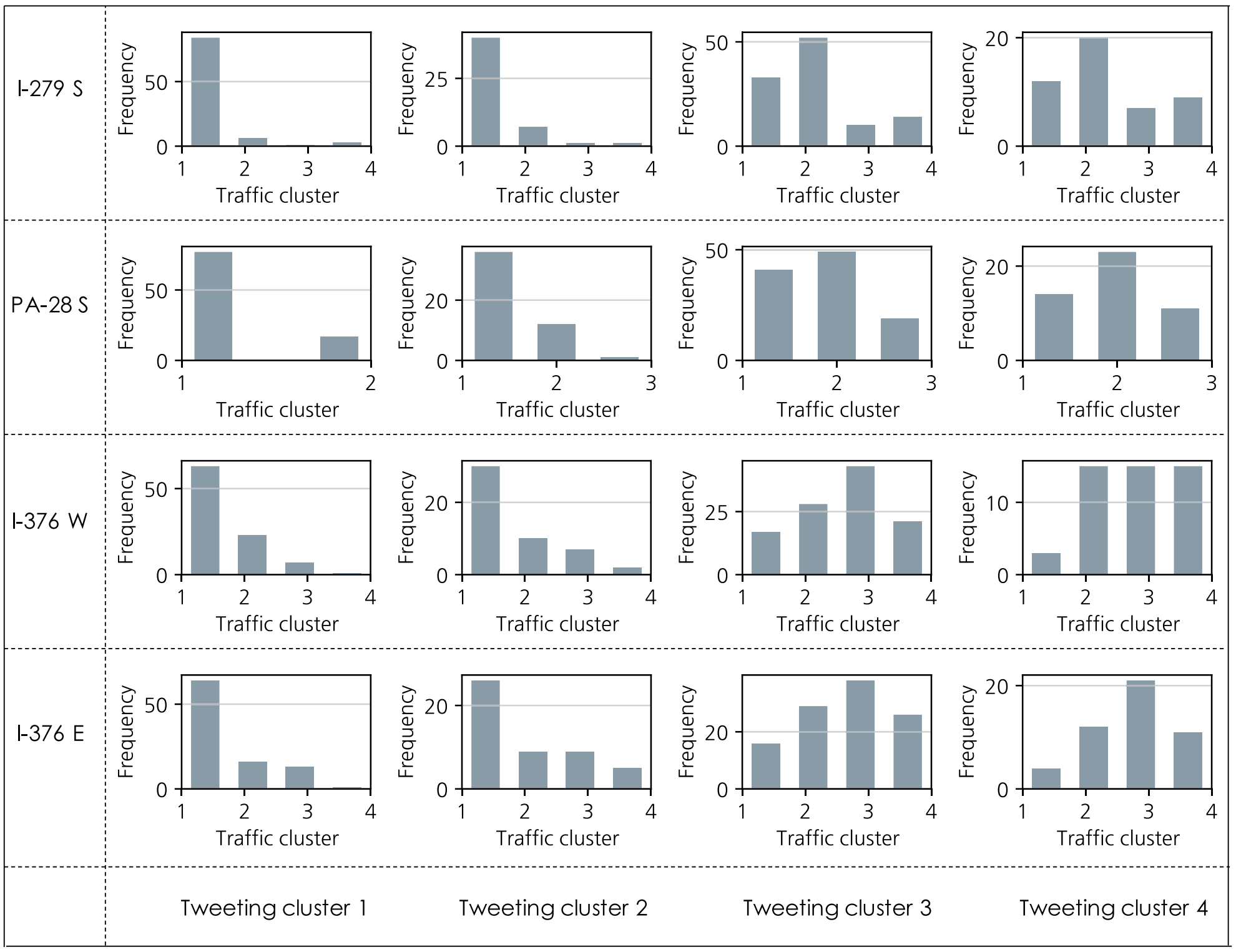

**Fig. 6.** Relationships between tweeting and morning traffic patterns.

**Table 2**
Chi-squared test and Cramer's V between traffic and tweeting clusters. Note: (1) Total sample size is 300 for all four roads, and (2) Cramers's V between 0.3 and 0.5 indicates moderate association.

| Road name | Number of traffic against tweet clusters | $\chi^2$ statistic | p-value | Cramer's V |
|---|---|---|---|---|
| I-279 S | 4 – 4 | 106.291 | <.001 | 0.330 |
| I-376 W | 4 – 4 | 110.003 | <.001 | 0.337 |
| I-376 E | 4 – 4 | 92.744 | <.001 | 0.307 |
| PA-28 S | 3 – 4 | 67.362 | <.001 | 0.321 |

Furthermore, we perform chi-squared test and compute Cramer's V for all four roads and present the results in Table 2. The chi-squared test indicates characteristic relationships between tweeting and traffic clusters with $p < .001$ for four roads in different spatial locations in Pittsburgh. Note that for simplicity of showing the relationship, we used discrete cluster representation and removed spatial and semantics information of tweets. While the Cramer's V is not high, the results have proved that tweeting activities explain some variances of next-day morning traffic. It is also important to note that the night period of tweeting activities before midnight is the key factor for explaining morning traffic.

By far, we show the great potential of mining Twitter messages for predicting next-day morning traffic. The moderate correlation calls for a model that effectively makes use of spatial and semantics information of tweets, along with supply-side features, such as weather and (planned) traffic incidents.

## 4. Method

Our proposed `tweet2traffic` predicts morning traffic, which is characterized by congestion status (CS), congestion starting time (CST), congestion duration (CD) and planning time index (PTI) for each of TMC segments in the regional network, using multi-source data provided before early morning. We first describe the data processing steps. The proposed model architecture is then presented. Finally, we introduce baseline models for benchmarking the prediction performances and model variants for showing the effectiveness of our design through ablation study.

### *4.1. Data processing*

Data processing steps for model outputs and inputs, which include traffic speed, social media, traffic incidents, and weather and temporal variables are introduced in this section.

#### *4.1.1. Morning traffic congestion output*

The morning period is defined as from 05:00 to 11:00 am, which consists of 72 data points of TMC speed measured every 5 min. Instead of predicting the entire morning speed time-series, we propose to describe morning traffic with a quadruple of variables $\mathbf{Y}_i^d = [CS_i^d, CST_i^d, CD_i^d, PTI_i^d]$, including congestion status (CS), congestion starting time (CST), congestion duration (CD) and planning time index (PTI). We argue that this model output representation is straightforward to predict and is sufficient as providing ahead-of-curve travel information to morning commuters. Our prediction goal is to predict these four congestion measurements before early morning (e.g. by 5am) or as late as midnight.

- Congestion status: a road segment $i$ is defined as congested at time $t$ on day $d$ ($CS_{it}^d = 1$), if the observed travel time index $TTI_{i,t}^d$, computed by Eq. (1). is greater than $TTI_{thres}$, i.e., $TTI_{i,t}^d \geqslant TTI_{thres}$, for at least $t_{min}$ minutes, as defined in Eq. (3). A segment $i$ is defined as congested on day $d$, i.e., $CS_i^d = 1$, if it is congested at any time during the morning period.

$$CS_{it}^d = \begin{cases} 1, \text{if } TTI_{it}^d, TTI_{it+1}^d, \ldots, TTI_{it+t_{min}}^d \geqslant TTI_{thres} \\ 0, \text{otherwise} \end{cases} \quad (3)$$

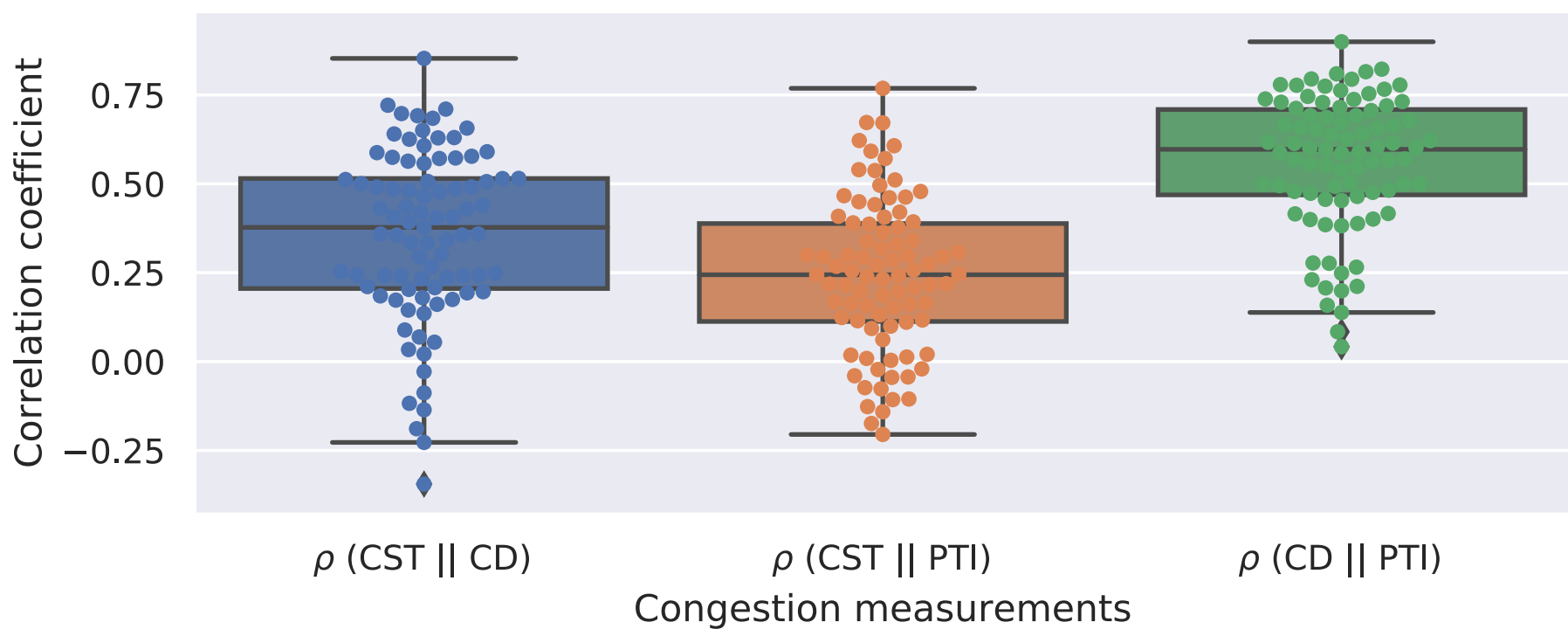

**Fig. 7.** Correlation between congestion measurements. Each data point represents the Pearson correlation coefficient between two chosen variables of a given road segment.

The term "**accuracy**" is used thereafter to indicate the percentage of binary congestion status that is predicted correctly.

- Congestion starting time: $CST_i^d$ denotes the starting point of the congested period, and if multiple congested periods exist, CST is the starting point of the earliest period. To be consistent, we reverse the time indices for computing CST, so that the earlier congestion occurs, the larger CST values are. For example, $CST = 72$ if congestion starts at 5am and $CST = 0$ if no congestion occurs during the morning.
- Congestion duration: $CD_i^d$ measures the length of congestion and is thus defined as the interval between the first congestion starting point and the last congestion ending point. The computation of congestion duration in this paper is related to how morning peak hours are generally defined. We argue if multiple congested periods exist and their time gaps are small (e.g., less than 15 min), the merged period better characterizes the morning peak hours. In addition, it is better to inform travelers of the start and end time of road congestion so they can avoid the entire morning rush hours.
- Planning time index: $PTI_i^d$ is computed as 95th percentile travel time during morning periods divided by free-flow travel time (Lyman and Bertini, 2008). It is a travel time reliability measure which represents how much total time a traveler should allow to ensure on-time arrival 95 percent of the time.

Note that congestion starting time, congestion duration and planning time index are only fit and predicted on congested days when congestion status $CS_{it}^d = 1$. For the remaining days, CST, CD and PTI are undefined. We further do correlation analysis to justify that these congestion measurements are not redundant. Initial results in Fig. 7 show correlations, i.e., $\rho(CD||PTI) > \rho(CST||CD) > \rho(CST||PTI)$, between the three congestion measurements. However, the Pearson correlation coefficients vary by road segments and are not strong enough for substitution between variables. More importantly, the measurements describe different aspects of morning traffic congestion. We argue that our defined congestion measurements, although entangled, are tailored to the needs of road users by providing interpretable and concise information to road users.

#### *4.1.2. Tweet processing pipeline*

We develop a social media processing pipeline to augment, clean, geocode and encode noisy Twitter messages into tweeting encoded feature vectors. We argue that tweets capture three types of information that can explain next-day morning traffic variances.

- Sleep-wake patterns: as illustrated in Fig. 6, tweeting activities in the night before and early morning capture the sleep-wake patterns in urban districts and thus explain next-day morning traffic;
- Event indicators: counts and sentiments of geocoded tweets, aggregated by space and time, have been proved in many prior works as an effective indicator for events and holidays, which can impact next-day morning traffic;
- (Planned) traffic incidents: Twitter accounts owned by public agencies and media (e.g., @511PAPittsburgh) automatically report real-time traffic incidents (e.g., crashes and planned roadworks, etc.), which have impacts on next-day morning traffic.

We process and encode separately the three types of information into features. The detailed steps are as follows.

*1. Augmenter.* A novel social media data augmentation method is proposed to alleviate user sparsity issues when extracting sleep-wake patterns in urban areas. Geocoded tweets are sparse because the sampled users change every day. For example, only 153 tweet users (1%) posted geocoded tweets between 9 pm-5am for more than 7 days in a year. The sparsity issue adds noises to the estimates of daily sleep-wake variances. We propose to reduce user sparsity by focusing on influential users, i.e., those who claim as residents in Pittsburgh and have posted a sufficient number of geocoded tweets that can be used to infer their approximate home locations. We use all their tweeting profiles scraped from user timelines retrieved by using the open-source Twint, which includes non-geocoded tweets, retweets, favorites, etc., to augment the estimates of sleep-wake pulses. Besides constructing a consistent user sample set, this augmentation method refines the estimates of sleep-wake pulses because (1) given that most residents sleep at home, the augmented tweets can recover full user tweeting activities in late night and early morning, while safely probing their geotags with home location; (2) to explain next-day traffic, spatially-segmented features extracted from tweets should focus on users who live in that community and thus commuting from that area next morning, rather than temporary visitors who posted geocoded tweets there last evening or night, but left afterward. The difficulty of this augmentation approach, however, is to filter influential users from a large number of noisy users (e.g. visitors, etc.), and to infer their home locations with geocoded tweets.

**Influential user filtering** Twitter users who posted at least five geocoded tweets in our dataset are selected. We then make use of self-reported user profiles to identify local residents of Pittsburgh. For users who have posted geocoded tweets within the bounding box, we select those who entered their residence with places in Pittsburgh. A resident classifier is carefully implemented using Regular Expressions (REs) that match local city names and nicknames (e.g. pittsburgh, pgh, da burgh, steel city, etc.), sports teams (steeler, etc.), zip codes (e.g. 15213, etc.), area code (e.g. 412), universities (e.g. cmu, chatham, etc.), townships (moon, robinson, etc.), neighborhoods (e.g. shadyside, oakland, etc.) and coordinates within the bounding box (e.g. 40.429, −79.932, etc.).

**Home location inference** We infer users' approximate home locations by two steps: (1) a density-based clustering algorithm called DBSCAN (Ester et al., 1996) is applied on tweet coordinates to identify a user's frequently visited places; (2) a rule-based classifier is tuned to find user homes from frequently visited places using features of that place cluster, including land-use variables, check-in ratios, night-time activities and home-related tweet features.

As illustrated in Fig. 8(a), DBSCAN requires two input parameters: *MinPts* and $\epsilon$. The algorithm iteratively starts with a random unvisited point in the graph and finds nearby points within $\epsilon$ radius neighborhood. A cluster is formed if more than *MinPts* points are

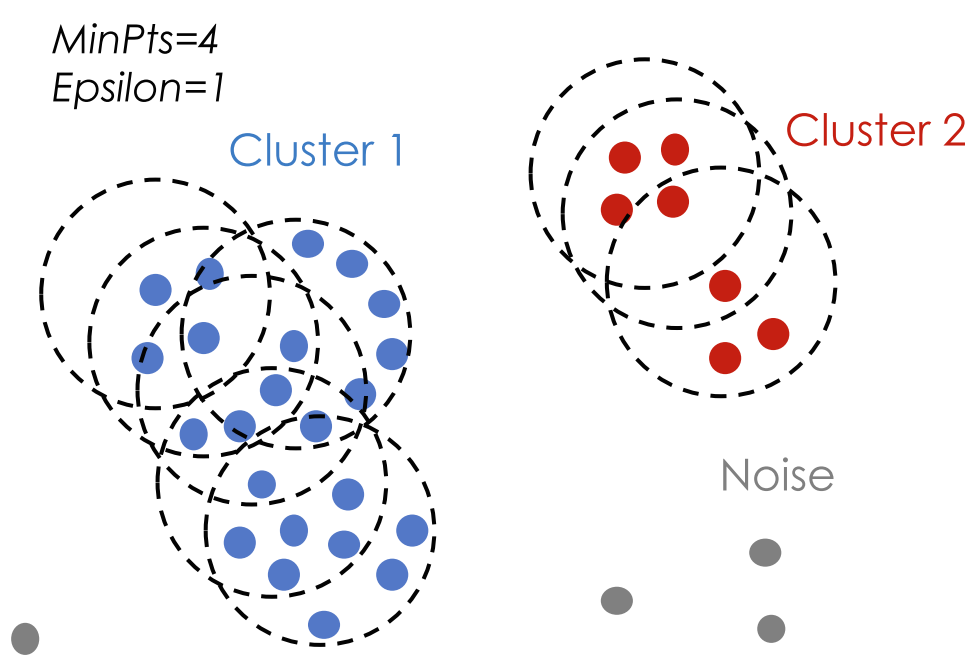

(a) DBSCAN clustering (Ester et al., 1996).

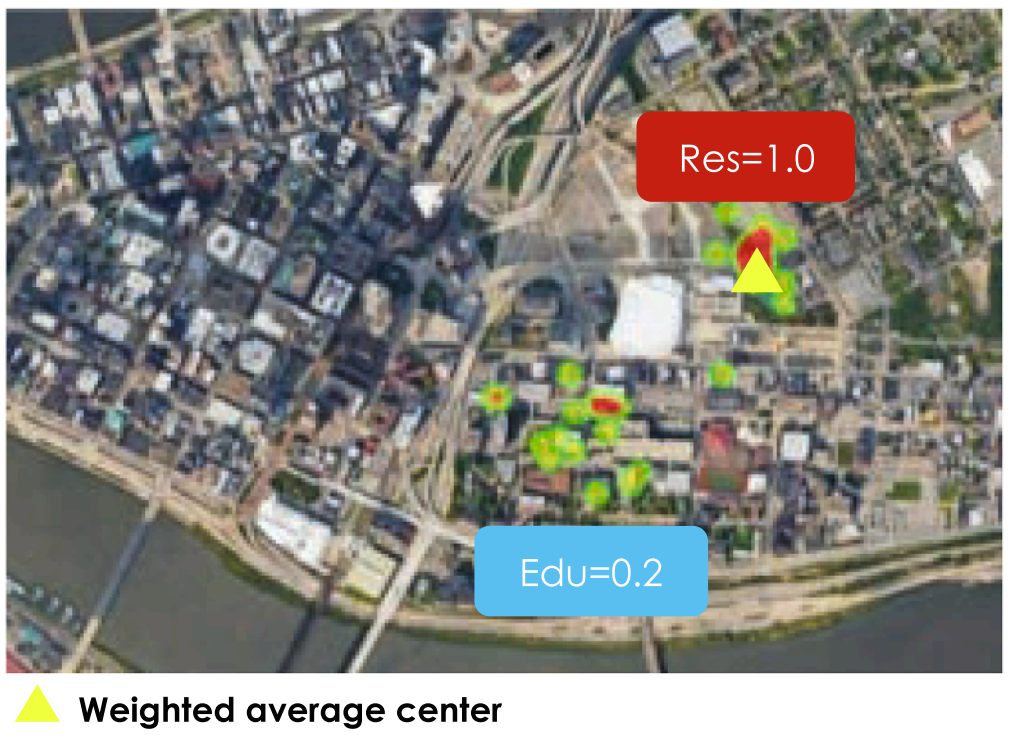

(b) Home clusters and weighted average home inference.

**Fig. 8.** Weighted user home location inference with DBSCAN.

found in the neighborhood area. The algorithm terminates if every point has been either visited or clustered and points not reachable from any cluster, at last, are identified as noise. Input parameters $\epsilon = 0.3$km and $MinPts = 1$ are used to cluster the coordinates of geocoded tweets by user. We set the radius $\epsilon$ to 0.3 km according to the average size of parcels in Allegheny County. *MinPts* was set to 1 such that an isolated check-in point for a user can form its own cluster.

We propose a rule-based classifier to find a user's home among identified clusters. For each of the coordinate clusters, we compute features including:

- Land-use composition: a vector containing lane-use composition (%) of tweets posted in that cluster. Pittsburgh zoning map[6] is used to join geocoded tweets spatially with land use information, which includes residence, downtown, education, industry, mixed-use, and amenity;
- Check-in ranking: a positive rank of tweet amount posted in that cluster by the user;
- Midnight activity: a binary variable indicating if the user has posted tweets in that cluster between 0 and 6am;
- Home tweets: a binary variable indicating if at least one home-related tweet is posted in that cluster. We apply a few keywords (e.g. sleep, wake, tv, sofa, bath, bed, etc.) to match tweets that users are likely to post at home;
- Last destination: a binary variable indicating if the cluster is the last destination of the user for at least one day.

Our proposed home classifier is an extension to the most-check-in method (Hossain et al., 2016). The classifier labels user home with six steps below:

1. `Initialize` all clusters as non-home cluster. `For` each influential user and `for` each of user's coordinate clusters, `repeat` the steps below in sequence;
2. `If` check-in ranking of the cluster is top 1, `and` midnight activity is positive, `and` industry and amenity account for less than 50% of land use, `then` label the cluster as user's home candidate;
3. `If` check-in ranking of the cluster is in top 3, `and` midnight activity is positive, `and` industry and amenity account for less than 50% of land use, `and` home tweets are positive, `then` label it as user's home candidate;
4. `If` that cluster is never user's last destination `or` industry and amenity account for more than 50% of land use, `then` label that cluster as non-home cluster;
5. `If` no home tweet has been posted in that cluster `and` other clusters of that user has been labelled as candidate home location, `then` label that cluster as non-home cluster;
6. `If` multiple clusters of a user are labelled as home candidates, the cluster with the highest check-in ranking are `then` chosen as home cluster. `Otherwise`, the home candidate is chosen as home cluster of that user.

After a user's home cluster is classified, check-in points within a home cluster can still spread across multiple land-use areas, as illustrated in Fig. 8(b). To locate user's home, handcrafted weights are used to approximate the probability of a point in that land-use area being a residential place, which include weights for: (1) residence = 1.0, (2) mixed-use = 0.5, (3) education = 0.2, (4) downtown = 0.2, (5) Industry = 0.0 and (6) amenity = 0.0. Finally, the coordinates of the user's geocoded tweets posted in the home cluster are weighted by land-use weights to approximate the user's home location.

**Geotagging** For influential users' timeline activities during late night and early morning (9 pm - 5am), we fill the missing tweeting coordinates with their inferred home locations by assuming residents sleep at home.

*2. Cleaner.* We develop two text cleaners separately for individual user tweets and traffic incident related tweets. Individual tweet

[6] Pittsburgh zoning map: https://gis.pittsburghpa.gov/pghzoning/.

cleaner is proposed for geocoded tweets and retrieved user timeline tweets, and traffic tweet cleaner is applied to traffic incident tweets posted by traffic agencies, i.e., @511PAPittsburgh in this study.

**Individual user tweets** Tweets contain spams and advertisements posted by bots, which should be removed for the purpose of extracting people's tweeting patterns. To identify these bots, we first filter suspicious accounts by computing their tweeting location ranges. 173 users with tweeting location range less than 10 meters are selected as suspicious bots in this step. Then, we adopt a publicly-available Botometer API (Davis et al., 2016) to give each suspicious account a "bot" score. Botometer is a random forest bot classifier built with more than 1,000 features extracted from available account meta-data, interaction patterns, and content scraped from 200 most recent tweets. Finally, 103 accounts with "bot" score higher than 2.0 are labeled as bots. The threshold is selected by manual inspection. After removing spams, we follow a common procedure described below to clean individual tweet content:

1. Lower the text content;
2. Remove special tweet entities (e.g. urls, emojis, email addresses, phone numbers, user names, etc.);
3. Segment hashtags to words and remove #, e.g., #LetsGoPens – lets go pens;
4. Concatenate consecutive (> 3) single-character tokens, e.g. Ain't H A P P Y – ain't happy;
5. Remove repeated suffix, e.g., Soooo goood lololol…– so good lol;
6. Translate slangs to formal words with a slang dictionary, e.g. lol – laughing out loud;
7. Remove special characters and brackets, e.g., *([)]&=;
8. Strip and remove extra whitespaces;
9. Add ending mark "." to unfinished sentences and fill empty tweets with an ending mark.

**Traffic incident related tweets** We extend the text processing steps in (Gu et al., 2016) to parse traffic incident tweets posted by public agencies. @511PAPittsburgh is an official Twitter account of the Pennsylvania Department of Transportation (PennDOT) for Southwest PA. The account reports real-time traffic incident status with a series of computer-generated tweets. For example, a full incident record is reported with tweets:

2019-12-27 06:42 – Multi vehicle crash on I-376 eastbound at Mile Post: 74.0. There is a lane restriction.
2019-12-27 07:18 – UPDATE: Multi vehicle crash on I-376 eastbound at Mile Post: 74.0. All lanes closed.
2019-12-27 08:02 – CLEARED: Multi vehicle crash on I-376 eastbound at Mile Post: 74.0.

Considering the fixed content format, we carefully implement Regular Expressions (REs) to extract highway or road names (e.g., I-376), direction (e.g., eastbound), exit or milepost (e.g., 74.0), incident type (e.g., crash), lane status (e.g., lane restriction – full closure – open) and tweet flag (e.g., occur – update – clear) from each incident tweet. Compared with (Gu et al., 2016), we add the retrieval of lane status and flags to better describe the incident record.

*3. Geocoder.* Individual (geocoded + timeline) tweets are spatially joined with census tracts on their posting coordinates. For traffic incident tweets, we applied the GIS developed in (Gu et al., 2016) to translate the parsed incident highway/road name and exit/mile post into incident latitude/longitude coordinates.

*4. Encoder.* Three types of information, i.e., sleep-wake patterns, local events and traffic incidents, embedded in tweets are encoded into features. Note that the three groups of features are extracted from different processed tweet datasets: (1) sleep-wake patterns are extracted from augmented influential user timeline tweets; (2) local event indicator features are built from geocoded tweets, and (3) traffic incidents are parsed from traffic incident tweets.

**Individual user tweets** Two groups of features are extracted from user tweeting activities and sentiment. To extract tweet

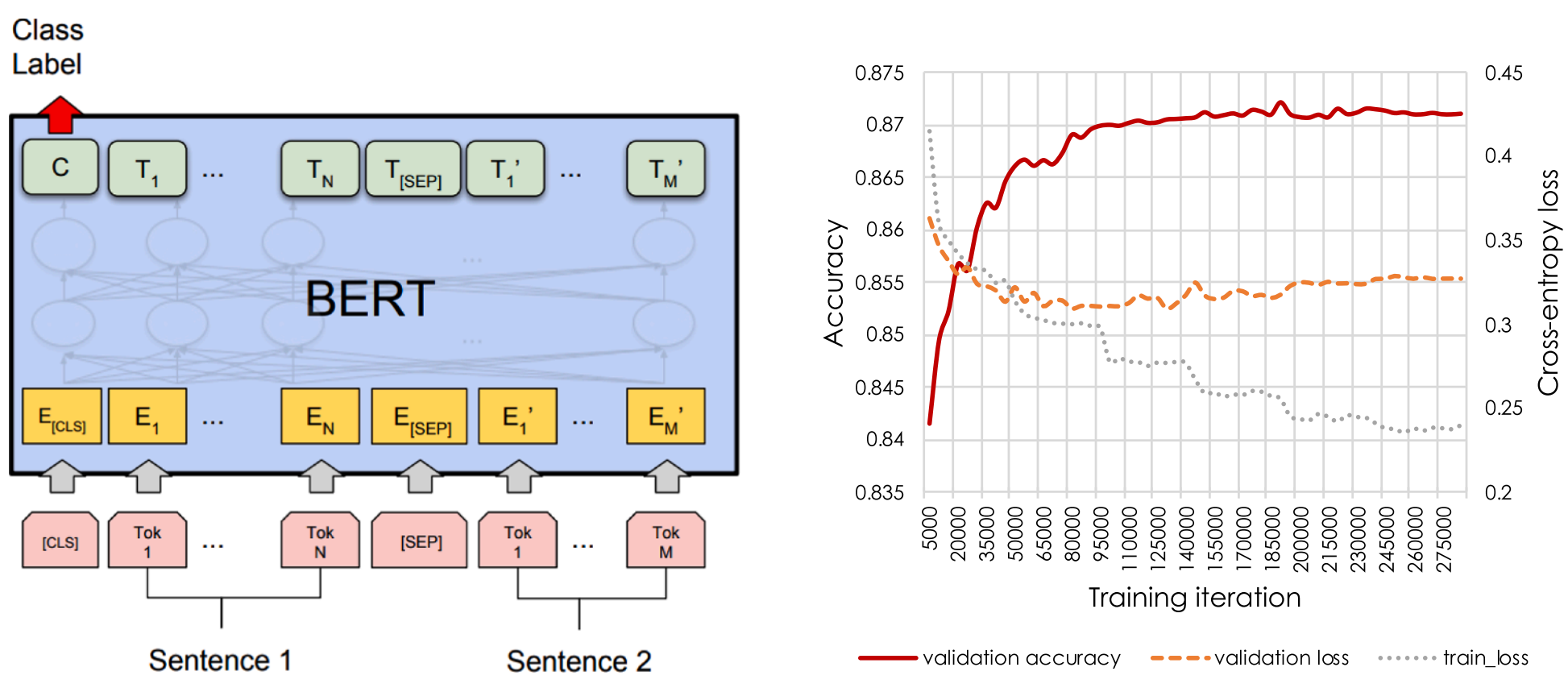

(a) BERT language model (Devlin et al., 2019). (b) Finetuning BERT on Sentiment140 dataset.

**Fig. 9.** BERT-based neural tweet sentiment model.

sentiment, a neural sentiment model is first constructed to embed tweet content into a vector. We finetune BERT (Devlin et al., 2019), a state-of-the-art neural language model shown in Fig. 9(a), on Sentiment140 dataset (Go et al., 2009). The dataset contains 1,600,000 tweets annotated with the polarity of content (i.e., Pos = positive, Neg = negative). We use the developed individual tweet cleaner to clean and normalize tweet content. A held-out dataset of 80,000 tweets is used to early-stop the training when validation accuracy does not increase. As shown in Fig. 9(b), the constructed model achieves an accuracy of 87.2% for classifying tweet sentiment on the validation dataset. Then, the fine tuned model is applied to user tweets to embed their content into vectors. We use the last sigmoid layer output $p$, i.e, the probability of a tweet being positive, to label content into three categories (Pos = positive, Neu = neutral, Neg = negative). We label tweet content as positive if $p \geqslant 0.7$, neutral if $0.3 < p < 0.7$ and negative if $p \leqslant 0.3$. The features extracted from user tweeting activities and sentiment are described below.

- Spatio-temporal sleep-wake activities: two histogram-like vectors describe the distributions of sleeping and waking up times of influential tweet users. The **last** augmented timeline tweets that the influential users post between 9 pm-3am and the **first** augmented timeline tweets they post between 3 and 5am are selected. Their counts, aggregated spatially by census tract and temporally by hour, are normalized by total counts to build the two histogram-like feature vectors. Instead of using keywords to match status that tweet users are likely to be in, this noisy feature labeling approach considers tweets without sleep-related content, requires less manual effort while proving effective in experiments for capturing overall sleep-wake patterns in the city.
- Event indicators: abnormal posting frequency and sentiment of **geocoded tweets** describe the occurrence of events. Six periods, including EM–Early Morning (3am-5am), AM–morning (5am-9am), DA–Mid-day (9am-6 pm), EV–Evening (6 pm- 9 pm), LN–Late Night (9 pm-0am), and MN–Mid Night (0am-3am) are used to segment the day. The number of geocoded tweets posted in these periods are used as features. Moreover, the percentage of neutral tweets in the six periods, among all sentiment categories, are also encoded as features. We expect that the tweeting frequency will be higher while the neural tweet percentage will be lower than average if some special events occur because most users either show positive or depressing attitudes towards events.

**Traffic incident related tweets** Traffic incident records, including closure/open timestamps, highway/road names, incident start/end coordinates, lane closure type are translated from the parsed incident tweet series. The same data format is also used in the PennDOT RCRS incident dataset so these two data sources can be easily integrated. Lane closure/open timestamps are defined as the first and last tweeting timestamps of the series of tweets describing an incident. If two or more exits/mileposts appear in a tweet, incident start location is defined as the coordinates of the smallest exit/milepost and incident end location uses the coordinates of the largest exit/milepost. Otherwise, the incident start and end locations both use the coordinates of the incident. Two types of lane closure, i.e., partial and full closure, are used to describe the severity of incident. We define an incident as a full-closure incident if lane status turns to full closure for at least once during the incident record.

#### *4.1.3. Traffic incidents*

Traffic incidents data, which include PennDOT RCRS incidents and the processed traffic incident tweets, are encoded so that those features represent the impacts of incidents on TMC segments. Three aspects of incident impacts are considered: (1) lane closure types, (2) incident location, and (3) incident time window. Two vectors, partial closure impacts $\mathbf{IF}_i^{L,P} = P_i(\mathbf{I}_i^L \otimes \mathbf{H}_i)$ and full closure impacts $\mathbf{IF}_i^{L,F} = F_i(\mathbf{I}_i^L \otimes \mathbf{H}_i)$, are extracted as follows.

*Impacts of lane closure type.* Let $P_i$ and $F_i$ be two complement binary variables describing the lane closure status of incident *i*. $P_i = 1$ when road is partially-closed and $F_i = 1$ when road has full closure. The two types of lane closure are encoded separately into two features $\mathbf{IF}_i^{L,F}$ and $\mathbf{IF}_i^{L,P}$, because lane closure types have been shown to impact very differently the change of traffic patterns in the network (Qian and Michael Zhang, 2012).

*Impacts of incident location.* To encode incident locations, a three-dimension vector $\mathbf{I}_i^L = [I_i^{DS}, I_i^C, I_i^{US}]$ is used. Fig. 10 shows the five possible incident locations relative to road segments. We use $L \in \{DS, C, US\}$ to denote locations, where $DS$ means the incident occurs on downstream, or intersects downstream of the TMC, $C$ means the incident contains the TMC, and $US$ means the incident is on upstream or intersects upstream of the TMC. Let $d_i$ be the minimum distance between incident start/end location and segment start/end location, and $d_{thres} = 5$ km is the farthest distance with which incident can impact the TMC in this study. Let $\mathbf{I}_i^L$ be the impacts of an incident $i$, which decays linearly with distance to incident location in Eq. (4). Note that elements in $\mathbf{I}_i^L$ range from 0 to 1.

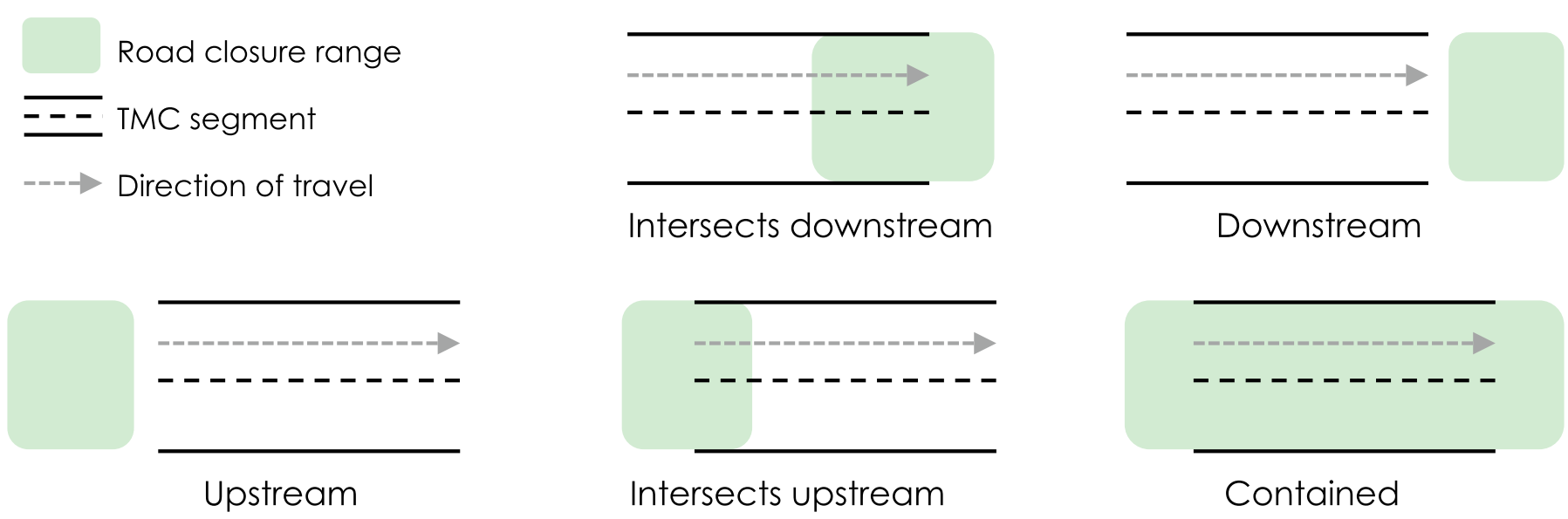

**Fig. 10.** Five types of road closure location w.r.t. road segment location.

$$I_i^{DS} = \max[0, (d_{thres} - d_i)/d_{thres}] \tag{4}$$

$$I_i^{US} = \max[0, (d_{thres} - d_i)/d_{thres}] \tag{5}$$

$$I_i^C = \begin{cases} 1, \text{if incident contains TMC} \\ 0, \text{otherwise} \end{cases} \tag{6}$$

*Impacts of incident time window.* A vector of binary variables $\mathbf{H}_i = [H_{i0}, \ldots, H_{iT}]$ is used where $H_{it} = 1$ if the road is closed in hour t for incident $i$. Note that $H_{i0}$ is included at the front to describe if there are incidents happening in midnight before 5am. Incidents that occur after morning periods, i.e., after 11am, are not considered.

#### 4.1.4. Weather and time features

Weather features used in this paper include six continuous variables – temperature, humidity, wind speed, pressure, visibility, hourly precipitation, and a binary variable – pavement condition. Continuous weather variables are scaled to 0–1 range by Min–Max normalization. Time features include four categorical variables: week-of-year, month-of-year, day-of-week, and holiday. We process them into an 8-dimension variable. For the cyclic month and week of year variables, we use sine and cosine functions to transform them into a two-dimension vector $[t_i^{(sin)}, t_i^{(cos)}]$:

$$t_i^{(sin)} = \sin(2\pi i/T) \tag{7}$$

$$t_i^{(cos)} = \cos(2\pi i/T) \tag{8}$$

where $i$ denotes the week/month index and $T$ denotes the total weeks/months in 2014. An advantage of this “clockwise” encoding is that each variable is mapped onto a circle such that the lowest value for that variable appears right next to the largest value (e.g. January is next to December). For day-of-week and holiday variables, we apply one-hot encoding after combining similar time features. Specifically, while Monday and Friday are encoded separately, Tuesday-Thursday are merged into one variable, so are Saturday, Sunday, and official holiday variables. To include prior and lagged effects of weekends and holidays, we include the number of days before and after the nearest holiday or weekend in the feature set.

### 4.2. Our proposed model

We develop a clustered learning pipeline that makes use of the spatio-temporal patterns of morning traffic along a road for predicting segment-level traffic. As shown in Fig. 11, the pipeline model consists of three building blocks: (1) road-level traffic descriptor, (2) segment-level traffic classifier, and (3) segment-level traffic regressor. By sharing the weights of road-level traffic descriptor for all segments along the road, spatial regularization is added to each segment-level traffic model by forcing them to predict the overall road congestion cluster index first.

#### 4.2.1. Road-level traffic descriptor

This module predicts morning congestion scales of a road. Because of spillback effects, morning road traffic shows ordered spatio-temporal clustering patterns as described in Section 3.1. It is natural to use ordered logit to model the relationship between road traffic cluster index and road-level input features, which include tweet features, weather and time variables. Note that traffic incident features are redundant at road level and hence are only included in segment-level models. We develop a set of “one-versus-rest” binary classifiers for the ordered logit model. The binary classifiers respectively predict if the road congestion cluster index is greater than a given value $c$. For example, the road-level descriptor of I-279 S which has four identified congestion clusters is comprised of three binary classifiers respectively predicting if the cluster index $c^d > 0, c^d > 1$ and $c^d > 2$. Due to the high dimensions and co-linearity of spatio-temporal tweet features, we apply $l_1$-norm regularization on model coefficients $\beta_c$ in Eq. (9) to remove irrelevant variables and to learn stable relationships.

$$\min_{\beta_c} - \sum_d \sum_{c=0}^{C-1} \left[ c^d > c \right] \log \left[ \sigma \left( \beta_c^T x_{\text{road}}^d \right) \right] + \lambda \|\beta_c\|_1 \tag{9}$$

Clustering features $\widehat{c}^d = \sigma(\beta_c^T x_{\text{road}}^d)$, which describe the predicted morning congestion scales of the road, are included in feature sets of segment-level models of the road.

#### 4.2.2. Segment-level traffic classifier

$L_1$-regularized logistic regression classifiers are separately built for each segment to predict next-day morning congestion status. As defined in Eqs. (10) and (11), $l_1$ penalty pushes the model to select critical features that explain segment-level variances, with road-level congestion explained by upper-level traffic descriptor $\widehat{c^d}$.

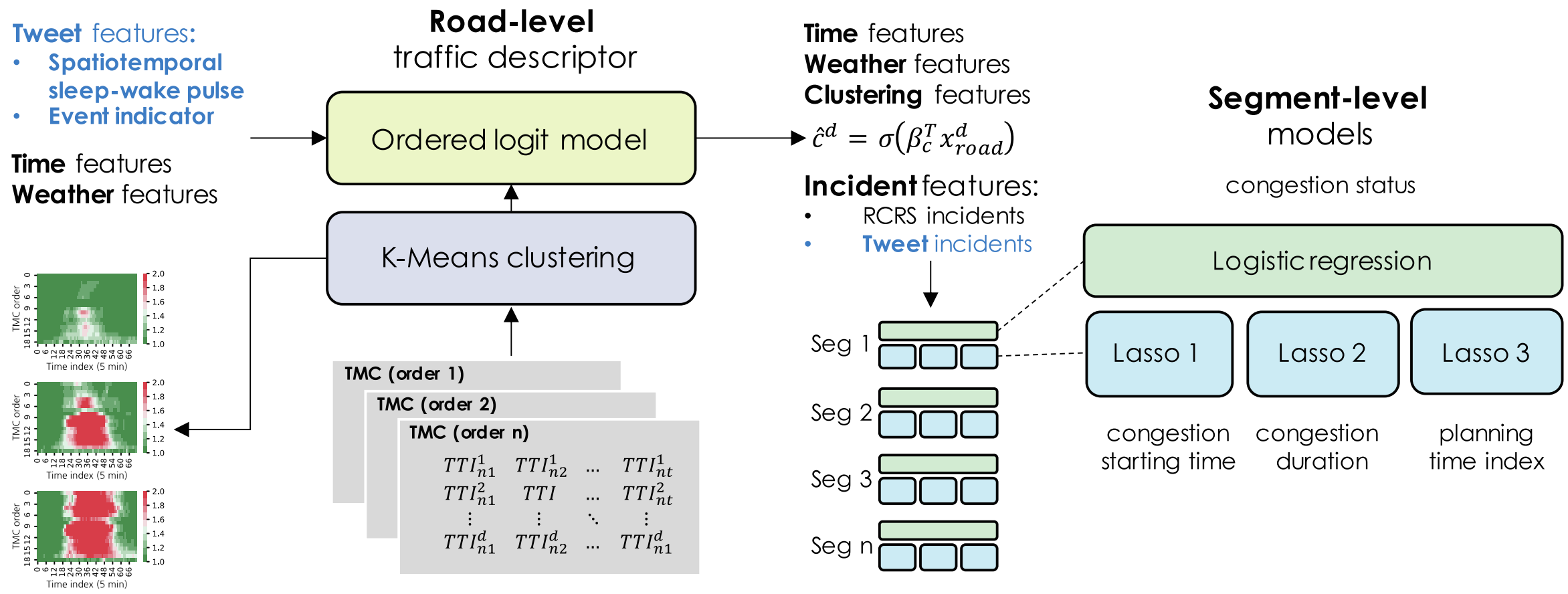

**Fig. 11.** `tweet2traffic` clustered model architecture.

$$\widehat{p}_i^d = \sigma(\beta_i^T[x_i^d, \widehat{c}^d]) \tag{10}$$

$$\min_{\beta_i} - \sum_d CS_i^d \log\widehat{p}_i^d + (1 - CS_i^d)\log(1 - \widehat{p}_i^d) + \alpha_i\|\beta_i\|_1 \tag{11}$$

For each segment, we fit two non-linear models on the selected features:

*Random-forest-classification (RF).* Random forest classification model trained on the features selected by $L_1$-regularized logistic regression.

*Neighbor-based-classification (KNN).* Neighbors-based classification model where the neighbors are obtained by selecting days with the $K$ closest learned road-level congestion scales $\widehat{c}^d$. Uniform weights are applied to traffic outputs of neighbor days to trigger predictions on congestion status $CS_i^d$.

### 4.2.3. Segment-level traffic regressor

Similarly, linear models with $l_1$ regularization (Lasso) are trained on days when congestion occurs ($CS_i^d = 1$) to predict congestion starting time ($CST_i^d$), congestion duration ($CD_i^d$) and planning time index ($PTI_i^d$) for each segment. These predictors learn weights $[\omega_i, \gamma_i, \eta_i]$ such that:

$$\min_{\omega_i}\|CST_i - \omega_i^T\left[x_i, \widehat{c}\right]\|_2^2 + \alpha_i\|\omega_i\|_1 \tag{12}$$

$$\min_{\gamma_i}\|CD_i - \gamma_i^T\left[x_i, \widehat{c}\right]\|_2^2 + \alpha_i\|\gamma_i\|_1 \tag{13}$$

$$\min_{\gamma_i}\|PTI_i - \eta_i^T\left[x_i, \widehat{c}\right]\|_2^2 + \alpha_i\|\eta_i\|_1. \tag{14}$$

For each segment, we fit two non-linear models on the selected features.

*Random-forest-regression (RF).* Random forest regression model trained on the features selected by Lasso.

*Neighbor-based-regression (KNN).* Similar to KNN classifier. Uniform weights are applied to traffic outputs of neighbor days to trigger continuous predictions on $[CST_i^d, CD_i^d, PTI_i^d]$.

## 4.3. Baselines and model variants

Two *baselines* are implemented: *HM* interprets recurrent traffic patterns with historical data, and *SAR* show the additional power of real-time traffic speed data before early morning for explaining morning peak hour traffic.

*Historical-mean (HM).* Baseline model for next-day morning traffic prediction without using real-time data. HM makes predictions using day-of-week historical average of traffic outputs $\mathbf{Y}_i^d = [CS_i^d, CST_i^d, CD_i^d, PTI_i^d]$ with a rolling window $H$ tuned by cross validation.

*Seasonal-autoregression (SAR).* Baseline model for traffic prediction with real-time traffic speed data. Let $P$ be the maximum non-seasonal lag and $H$ be the maximum seasonal lag. As defined by Eq. (15), the non-seasonal terms, i.e., real-time traffic speed before early morning $v_{it-p}^d$ and seasonal terms, i.e., past day-of-week seasonal speeds $v_{it}^{d-7h}$ are combined to predict TMC speed in the morning through an autoregressive way. $w_p$, $W_h$ and $c$ are learnable model weights. Traffic outputs $\widehat{\mathbf{Y}}_i^d$ are computed from speed predictions $\widehat{v}_{i,t}^d$ following data processing steps in Section 4.1.1.

$$\hat{v}_{it}^{d} = c + \sum_{p=1}^{P} w_p v_{it-p}^{d} + \sum_{h=1}^{H} W_h v_{it}^{d-7h} \quad (15)$$

We experiment with three *model variants* to show the effectiveness of our model design.

*Tweet2traffic-no-feature.* A model variant trained with the removal of a selected feature type. We assess the role of each data source by removing its all processed features from our model and computing the performance deduction.

*Tweet2traffic-no-cluster.* A model variant without road-level traffic descriptor (i.e., k-means clustering and ordered logit model blocks). The high-dimensional social media features are fed into segment-level models directly. Other blocks (e.g. classifier, regressor, regularization, etc.) remain the same.

*Tweet2traffic-before-time.* A model variant trained with data (e.g. tweet, weather, traffic incidents, speed, etc.) accessed before an earlier *time* than 5am. This variant tests how ahead-of-curve our model can predict morning traffic, by quantifying the trade-offs between forecasting horizon and prediction performances.

## 5. Results and discussion

We run experiments with 300 days of traffic and social media data on 53 TMC road segments in the City of Pittsburgh in 2014. For all experiments, we use $TTI_{thres} = 2$ and $t_{min} = 15$ min to identify morning congestion and to generate ground-truth traffic congestion labels. For days with multiple congested periods, we combine the periods if their gap is less than 15 min. We've used different training and testing datasets for (1) evaluating prediction performances and (2) interpreting the model, which are described as follows:

1. When evaluating the model prediction performances, our model uses data known before 5am, or earlier discussed in Section 5.4 to inform morning traffic predictions of $\mathbf{Y}_i^d = [CS_i^d, CST_i^d, CD_i^d, PTI_i^d]$ on all segments, described by congestion status ($CS_i^d$), congestion starting time ($CST_i^d$), congestion duration ($CD_i^d$) and planning time index ($PTI_i^d$) in morning periods. Hence, for traffic incidents, only planned traffic events (e.g. road construction, bridge precaution, etc.) and crashes reported before 5am (or earlier depending on the cut-off time of data feed) are fed as inputs to the model. To train the models and evaluate performances, nested time-series cross validation (tsCV) is employed. In the outer loop, tsCV splits the dataset in sequence such that in the kth split, the first k folds are used for training and the (k + 1)th fold is for testing. The inner loop applies cross validation (CV) to tune hyperparameters. Finally, the model is fitted on the whole training set with tuned model hyperparameters and evaluated on the test set in the outer loop. The experiment setting ensures that: (1) model training and hyperparameter tuning access only training sets, and (2) future information is not abused to predict the past when testing. The models are evaluated on each testing split by metrics including classification accuracy, precision and recall score for $CS_i^d$, and root-mean-squared-error (RMSE) for $CST_i^d$, $CD_i^d$ and $PTI_i^d$. Note that for classification of $CS_i^d$, all training and testing samples are used by models but for regression of $[CST_i^d, CD_i^d, PTI_i^d]$, only days with morning congestion (i.e., $CS_i^d = 1$) are fed into the model. Similarly, models are only tested on congested days. By applying 10-fold tsCV in the outer loop and 4-fold CV in the inner loop, we report the model prediction performances by each model and testing split. An aggregated score weighted by the number of samples in each testing split is also reported for each model.
2. When interpreting model parameters and predictions, all 300 days of data are used for training, and hyperparameters are selected by 4-fold cross validation. This is to ensure the fitted model parameters are consistent across the splits. All types of traffic incidents (i.e., planned and unplanned traffic events) in the morning are assumed to be known to the model.

**Table 3**
List of features used by the predictive model.

| | |
|---|---|
| | **Tweet features** |
| Spatio-temporal sleep-wake pulse | Aggregated count distribution of influential users' last tweets in night and first tweets in early morning by hour and census tract (Feature name: *Hour_CensusID*) |
| Aggregated count | Aggregated geocoded tweet count distribution by day period defined in Section 4.1.2 (Feature name: *Period*) |
| Aggregated sentiment | Percentage of neutral geocoded tweets by day period (Feature name: *Neu_Period*) |
| | **Incident features** |
| Partial closure impacts ($IF_{it}^{L,P}$) | Traffic impacts of a partial closure incident at location $L$ of the road segment in hour $t$ (Feature name: *p_Location_Hour*) |
| Full closure impacts ($IF_{it}^{L,F}$) | Traffic impacts of a full closure incident at location $L$ of the road segment in hour $t$ (Feature name: *f_Location_Hour*) |
| | **Weather features** |
| Air temperature | Air temperature in degree F immediately by hour (Feature name: *temp_Hour*) |
| Relative humidity | Relative humidity in percent by hour (Feature name: *hum_Hour*) |
| Wind speed | Wind speed in feet by hour (Feature name: *wspd_Hour*) |
| Pressure | Mean sea level pressure by hour (Feature name: *pressure_Hour*) |
| Visibility | Prevailing hourly visibility (Feature name: *vis_Hour*) |
| Pavement condition | Binary variable (1 for wet, icy, flooded) by hour (Feature name: *pave_cond_Hour*) |
| Adverse weather | Categorical descriptions of observed adverse weather conditions. Larger values indicate severer weather conditions (Feature name: *wx_phrase_Hour*) |
| Hourly precipitation | Hourly precipitation (Feature name: *precip_hrly_Hour*) |
| | **Time features** |
| Cyclic month | Clockwise encoding of month of year (Feature name: *mon_sin, mon_cos*) |
| Cyclic week | Clockwise encoding of week of year (Feature name: *week_sin, week_cos*) |
| Holiday | Binary variable indicating if the day is a holiday or during weekend (Feature name: *dow_wkd_holiday*) |
| Day of week (Monday) | Binary variable indicating if the day is Monday (Feature name: *dow_mon*) |
| Day of week (Tues–Thurs) | Binary variable indicating if the day is between Tuesday and Thursday (Feature name: *dow_tue_thu*) |
| Day of week (Friday) | Binary variable indicating if day is Friday (Feature name: *dow_fri*) |
| Prior holiday effects | Number of days before next holiday or weekend (Feature name: *nxt_rest*) |
| Lagged holiday effects | Number of days after last holiday or weekend (Feature name: *lst_rest*) |
| | **Clustering features** |
| Road congestion scale | Probabilities of traffic congestion on road greater than a given level predicted by road traffic descriptor (Feature name: *c_Level*) |

The features used by the predictive model are summarized in Table 3. We compare model performances with baselines in Section 5.1. Section 5.2 elaborates the respective contributions of tweets, traffic events and weather data to model prediction, what data patterns it captures, and discusses when and where our method outperforms the variants. In Section 5.3 we show the effectiveness of our proposed clustered model structure. How in advance can morning traffic be predicted with social media data is further examined in Section 5.4.

### 5.1. Next-day morning traffic prediction

The model prediction performances of our proposed model `tweet2traffic` against baselines, described by aggregated evaluation metrics across testing splits, are summarized in Table 4 and visualized by road segment in Fig. 12. The poor prediction performances of two baseline models, i.e., *seasonal-autoregression (SAR)* and *historical-mean (HM)*, confirm the limitation of using real-time speed data for morning traffic prediction in Section 1 generally exist in traffic networks and Fig. 1 is no special case.

First we find that the autoregression-based method (SAR) can trigger reasonable traffic predictions of the next time step (5 min) for most road segments ($R^2 = 0.83 \pm 0.10$). However, when used to predict traffic for whole morning periods using data by 5am, SAR can often predict no congestion at all, which is supported by the extremely low recall of congestion status and high prediction errors of congestion starting time, duration and planning time index presented in Table 4. This is because congestion often doesn't exist or traffic demand can grow very slowly prior to 5am on some weekdays, and SAR is unable to pick up the drastic traffic break-down using only the traffic data up to 5am. In addition, adding the seasonal features to AR models doesn't seem to be a solution for the problem setting, as the model puts most weights on real-time traffic speed (i.e., AR(1) term), which can only be inferred autoregressively from model predictions if using only the traffic data up to 5am. On the contrary, *historical-mean (HM)* baseline which doesn't use any real-time information performs much better than SAR in terms of predictions of all congestion measurements. This is no surprise, as HM offers insights for day-to-day recurrent traffic. This is why HM is used by many traffic management centers to tune next-day traffic management strategies beforehand, and is offered as the 'typical traffic' in Google Maps'. However, it is shown in Table 4 that the overall prediction performances of HM, with 79% accuracy for predicting congestion status, 48 min error for congestion starting time and 74 min error for congestion duration, do not provide adequate travel advice for morning commuters.

Results show that our proposed models `tweet2traffic` generally outperform baselines for all congestion measurements and the performance improvement is more significant for predictions of congestion status and congestion starting time, if compared with the HM baseline. In addition, if compared with two nonlinear model variants of `tweet2traffic` using random forest (RF) and neighbors-based (KNN), the model performances degrade on most predictions of congestion measurements. Although one may expect adding feature nonlinearity often boosts performances, our experiments show otherwise, possibly because of the relatively small sample size in some time-series splits. To further locate the sources of performance improvement, the percentage improvement of our method is

**Table 4**
Comparisons of model performances averaged over evaluation periods and road segments. The reported standard deviations (±) account for variances of evaluation metrics across different road segments.

| **Method** | **Congestion measurements** | | | | | |
|---|---|---|---|---|---|---|
| | Congestion status (%) | | | Starting time (h) | Duration (h) | Planning time index |
| | Accuracy | Precision | Recall | RMSE | RMSE | RMSE |
| **Our model** | | | | | | |
| Tweet2traffic | 0.88 ± 0.05 | 0.79 ± 0.16 | 0.85 ± 0.15 | 0.57 ± 0.22 | 1.05 ± 0.76 | 6.53 ± 3.98 |
| Tweet2traffic-RF | 0.86 ± 0.06 | 0.77 ± 0.16 | 0.78 ± 0.19 | 0.65 ± 0.28 | 1.16 ± 0.88 | 7.12 ± 3.87 |
| Tweet2traffic-KNN | 0.85 ± 0.07 | 0.78 ± 0.14 | 0.75 ± 0.20 | 0.61 ± 0.25 | 1.19 ± 1.01 | 6.56 ± 3.72 |
| **Baseline** | | | | | | |
| SAR | 0.58 ± 0.16 | 0.22 ± 0.29 | 0.09 ± 0.20 | 3.50 ± 0.59 | 2.33 ± 1.57 | 15.29 ± 13.28 |
| HM | 0.79 ± 0.07 | 0.60 ± 0.24 | 0.68 ± 0.26 | 0.80 ± 0.34 | 1.24 ± 0.78 | 8.15 ± 5.17 |

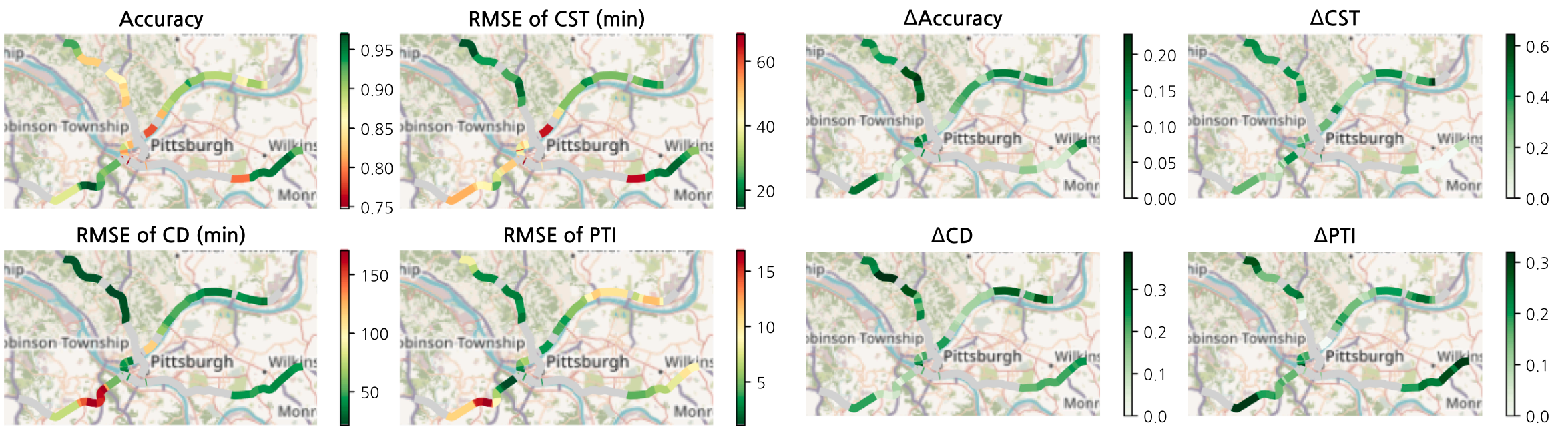

(a) Morning traffic prediction error by road segment.

(b) Performance improvement against historical-mean baseline.

**Fig. 12.** Visualization of `tweet2traffic` congestion prediction model performances by road segment: (a) Congestion prediction error of our model by road segment, and (b) percentage improvement against historical-mean baseline. Note: RMSE = root mean square error; CST = congestion starting time; CD = congestion duration; PTI = planning time index; Δ = percentage improvement.

visualized against HM by road segment in Fig. 12.

It is found that the model prediction performances differ a lot by spatial locations. As shown in Fig. 12(a), some segment models can offer accurate CST predictions with less than 20 min error and CD predictions with less than 25 min error, while some other segments can have CST error over 65 min and CD error even over 2 h where no significant improvement over HM is observed. A similar pattern is observed for percentage performance improvement against HM baselines in Fig. 12(b), where `tweet2traffic` can have 64% CST prediction improvement and 39% CD improvement for some special locations (e.g. at the edge of spillback regions upstream to traffic bottlenecks) in the regional network, while several segments only show a trivial improvement in terms of all congestion measurements with multi-source data features added. Besides different traffic conditions, the results suggest that the multi-source data features, including tweets, incidents, weather and temporal variables, are only effective for improving morning traffic predictions on some locations in the network, but not for all. This is completely anticipated as some road segments are much more likely to be influenced or implied by weather, population activities and incidents than others. The data driven approach precisely helps learn the "signals" of morning traffic congestion as much as it can from social media data, which is effective and powerful on a portion of roads. The `tweet2traffic` model may not work for other roads, simply because social media are not actively used in areas that are relevant to those roads.

It is worth noting that the `tweet2traffic` model works the best on those segments that are upstream of "choking' points", namely spill-back segments of bottlenecks with large day-to-day congestion variations. This is particularly appealing since this next-day traffic information can be the most valuable to both commuters and operators.

## *5.2. Role of each data source*

We explain how multi-source data impact overall morning traffic prediction in Section 5.2.1. The respective contributions of each data source, and when and where our model outperforms those variants is discussed in Section 5.2.2.

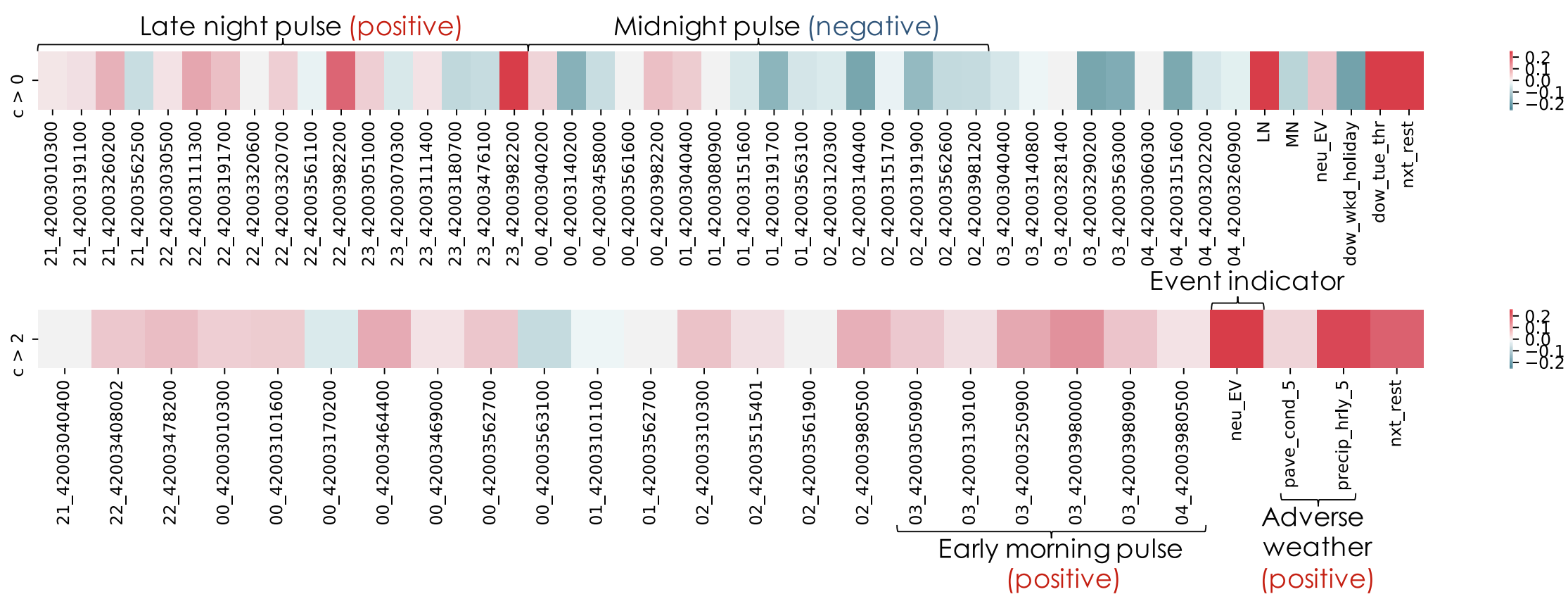

**Fig. 13.** Model weights of road-level traffic descriptor of I-279 Southbound. Note: $c$ = congestion cluster index; full names and details of features used by the model can be found in Table 3.

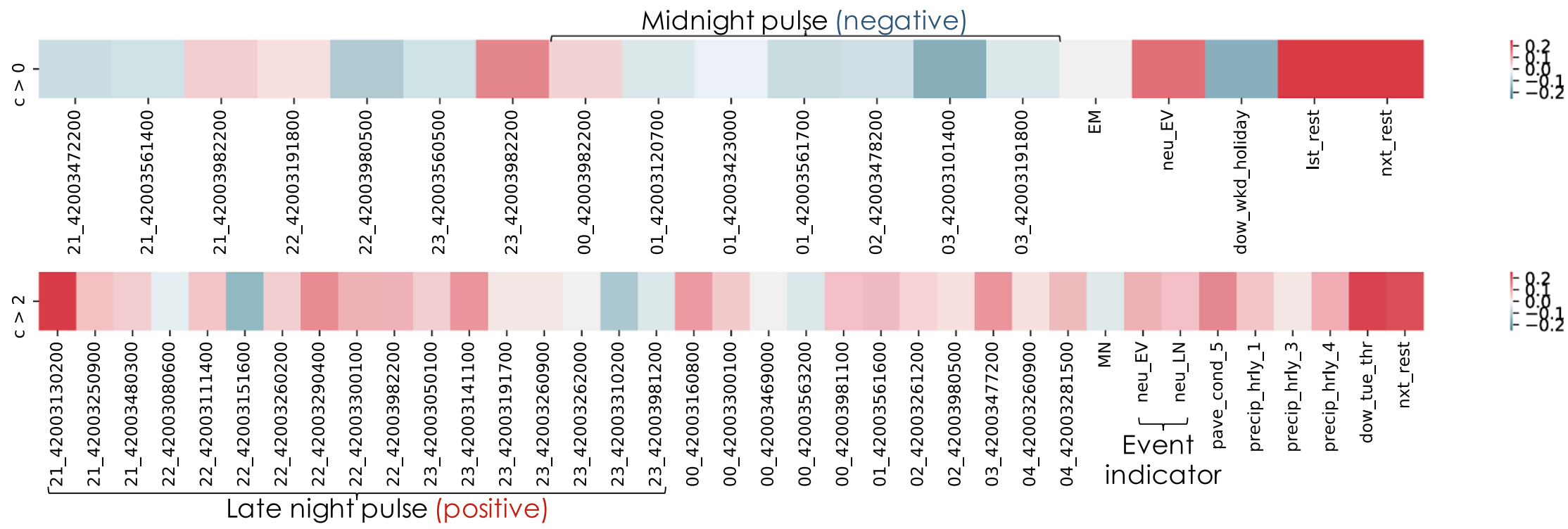

**Fig. 14.** Model weights of road-level traffic descriptor on I-376 Westbound. Note: $c$ = congestion cluster index; full names and details of features used by the model can be found in Table 3.

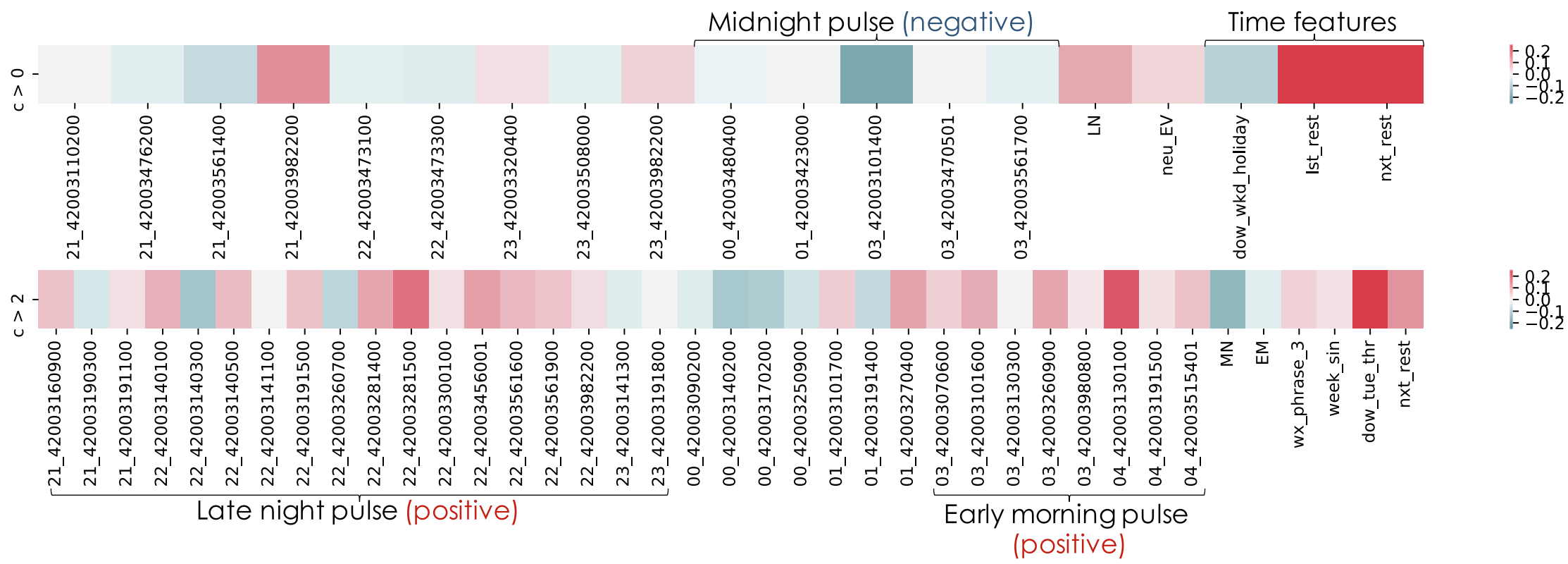

**Fig. 15.** Model weights of road-level traffic descriptor on I-376 Eastbound. Note: $c$ = congestion cluster index; full names and details of features used by the model can be found in Table 3.

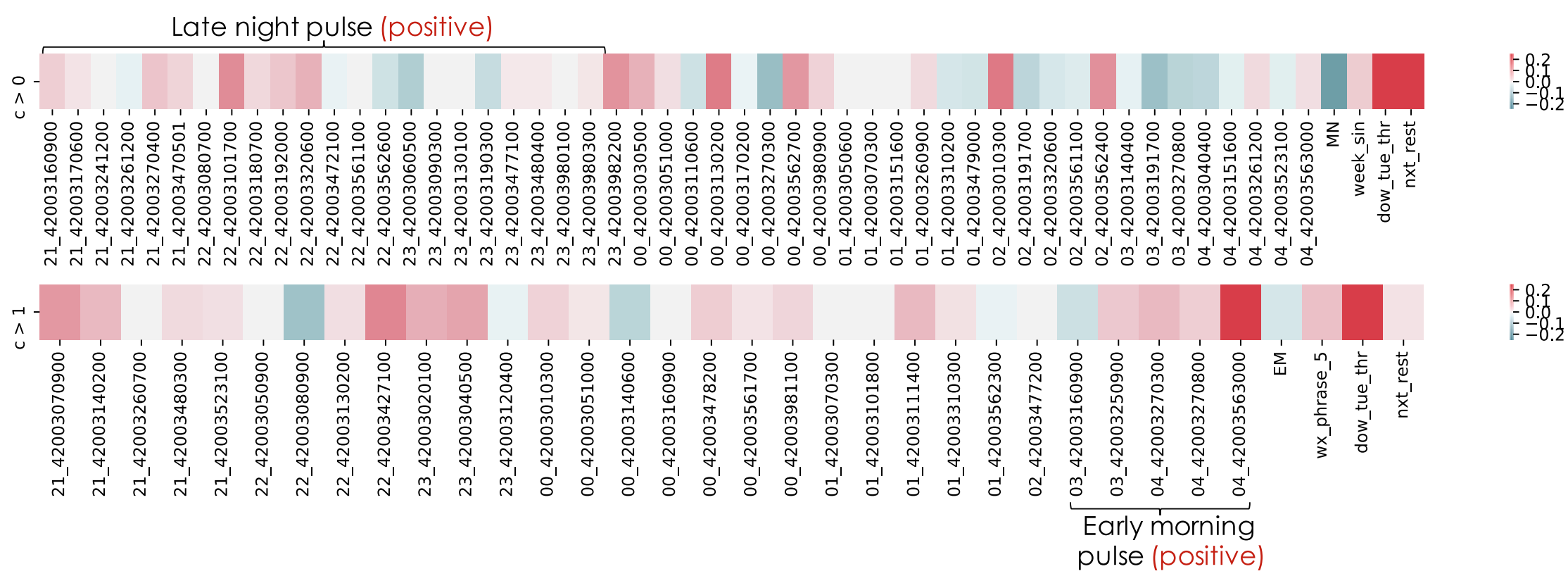

**Fig. 16.** Model weights of road-level traffic descriptor on PA-28 Southbound. Note: $c$ = congestion cluster index; full names and details of features used by the model can be found in Table 3.

#### 5.2.1. *Visualization of model parameters for road-level traffic descriptor*

We first visualize in Figs. 13–16 the weights of road-level traffic descriptors fitted on four roads, i.e, I-279 S, I-376 W, I-376 E and PA-28 S used in our study. The road-level models describe the overall congestion patterns on the road by predicting if the road congestion cluster index $c$ is larger than a given level. For example, the road-level descriptor of I-279 S which has four identified congestion clusters (Fig. 4) is comprised of three binary classifiers respectively predicting if $c > 0, c > 1$ and $c > 2$. For interpretability, the model weights of the first and last classifier of each road-level traffic descriptor, which are defined in Eq. (9), are presented. The weights of the descriptor explain how multi-source tweets, weather and temporal features impact the morning traffic at the road level. The positive (red) model weights imply positive impacts on morning traffic congestion while negative (blue) model weights generate negative impacts. The impacts of traffic incident information will be examined by the segment models in Section 5.2.2.

Clearly, the selected features are consistent with our anticipation. As shown in Figs. 13–16, the selected features are across all types of tweet features, weather and temporal variables. It is found that tweeting features, which include both spatiotemporal sleep-wake pulses and event indicators, serve as a proxy for morning travel demand. As expected, the spatiotemporal sleep-wake pulses captured by each descriptor differ by space (i.e., census tract) and time (i.e., hour index) as morning travel demand among different Origin–Destination pairs is loaded into these four roads by different time. However, when looking at each road descriptor separately, the resulting relationships are surprisingly simple. Generally, we discover that the earlier people go to sleep, the more congested the road will be in the next morning. The early-sleeping pulse is represented by high resident tweeting activities in the evening, together with low tweeting activities during midnight in selected spatial areas. Accordingly, it is found in Figs. 13–16 that weights of sleep-wake pulses between 9 pm and midnight are mostly positive for $c > 0$ classifiers while weights of sleep-wake pulses between 0-3am are negative. This supports our finding that early-sleeping pulse positively impacts morning congestion if the roads will be congested the next morning. Interestingly, we also find that the early-wake pulses, represented by high resident tweeting activities in the early morning, also positively relates to the level of morning congestion on the road. This finding is supported by the positive weights of sleep-wake pulses between 3-5am for $c > 1, 2$ classifiers in Figs. 13–16. It strongly supports our anticipation that in general the earlier people go to sleep, the higher chances they will commute and depart early, both of which contribute to morning congestion.

Next, we present the impacts of tweet event indicator features which comprise of aggregated geocoded tweet counts and percentage of neutral tweets by time period. It is found that the percentage of neutral tweets in the evening (6–9 pm), i.e., labeled as "neu_EV", is positively-correlated with next-day morning road congestion. This suggests that the occurrence or discussion of big events in the evening, represented by higher or lower tweet sentiment than normal, often results in low travel demand in the next morning. Possible examples of "big events" include the occurrence of sports games (e.g., Pittsburgh Pirates, Penguins, Steelers, etc.) in the evening or discussion of upcoming local holidays and so on, which are reasonable indicators of low travel demand in next morning.

As expected, roads are found to be more congested between Tuesday and Thursday and traffic during weekends and official holidays are of low volume. Surprisingly, we show that weather features such as precipitation, pavement conditions and adverse weather conditions are only selected in $c > 1, 2$ classifiers. This suggests that adverse weather conditions in the early morning can often impact the morning road congestion levels if congestion occurs, but traffic congestion may be completely avoided when commuters become aware of the adverse weather in advance in the next morning.

### 5.2.2. *Ablation study of data source*

Experiments with `tweet2traffic-no-feature` model variants are conducted to determine the contribution of each data source to model prediction performances. We remove specific features from the model and computes performance degradation from `tweet2traffic`. Note that we only evaluate the importance of data sources on the top 10 road segments selected by performance improvement against *historical-mean* baseline in Section 5.1, as these segments make the most use of multi-source data for traffic prediction. Note that most of these segments are located at the edge of spillback regions upstream to traffic bottlenecks in the regional network. The results are shown in Fig. 17.

It is surprising to see that if we do not include Twitter related features, the model performances drop the most on all selected road segments in terms of all four congestion measurements. As shown in Fig. 17, the prediction of congestion duration and planning time index degrades over 15% if not including tweet features. On the contrary, if we remove traffic incident or weather features, the performance only degrades by 4% and 6%, or 5% and 4% for CD and PTI. This finding highlights the importance of having social media data for some specific road segments. There are three main reasons: (1) only traffic events and weather conditions reported before 5am are used to fit the models so a large portion of traffic incidents, such as crashes during the morning commute, cannot be used by any models. The same applies to weather conditions, where moderately adverse weather conditions during the morning commute that do not prevent commuting trips can have unknown effects to traffic; (2) social media related features can be moderately correlated with those weather and indent related features. In other words, tweets may already include some information related to weather and incidents; and (3) traffic events and adverse weather conditions are sparsely distributed in time and space so the overall performance may not be affected significantly. Those hypotheses call for an investigation into testing samples to examine when and where our model outperforms `tweet2traffic-no-feature` variants.

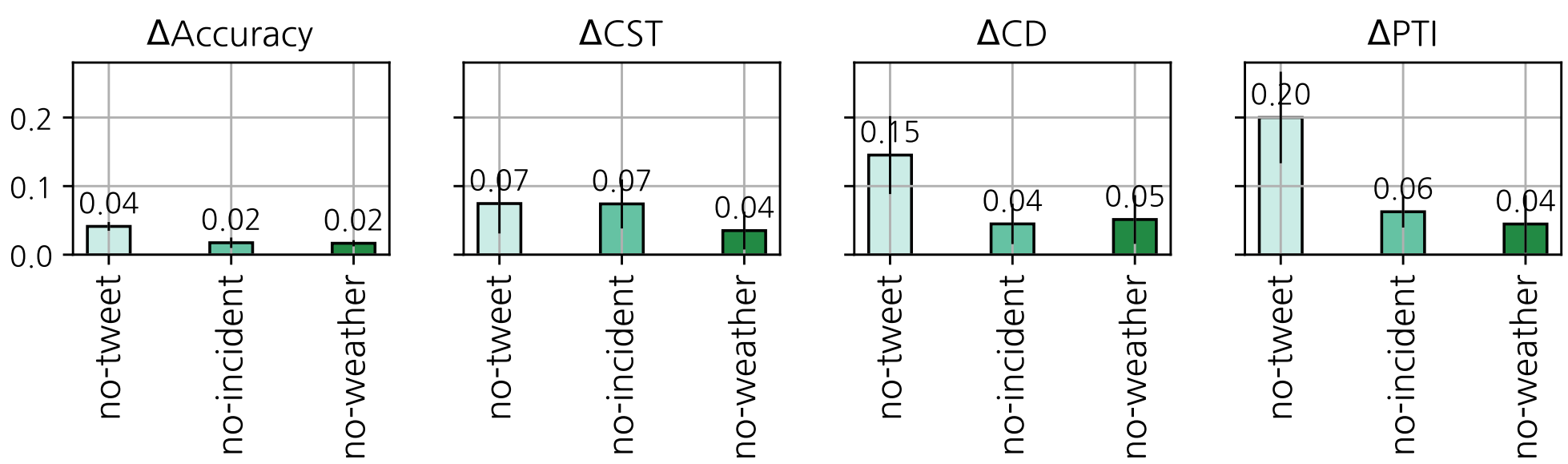

**Fig. 17.** Percentage performance degradation by removing data source for the top 10 road segments. The error bars account for variances across road segment. Note: CST = congestion starting time; CD = congestion duration, and PTI = planning time index.

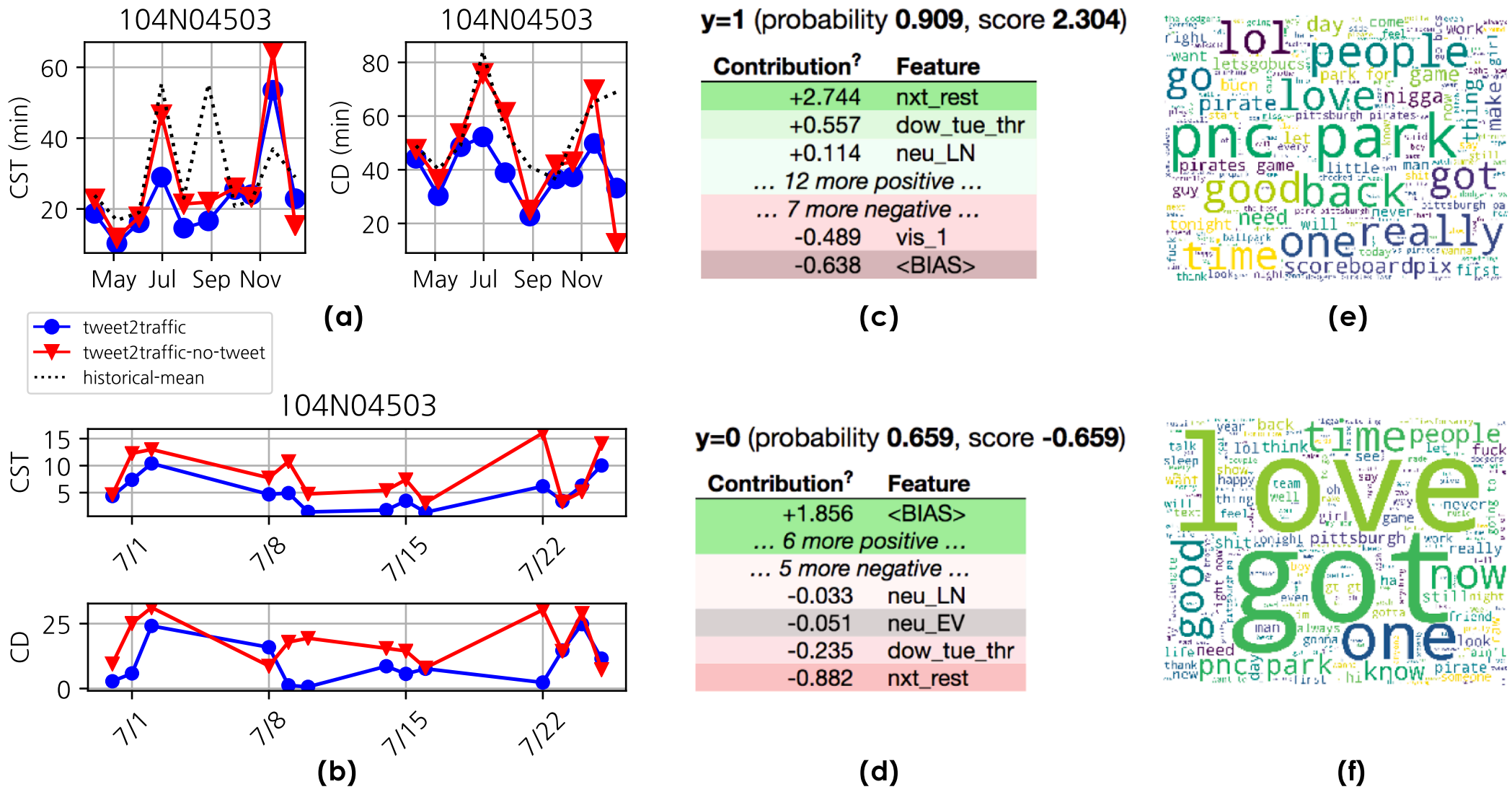

**Fig. 18.** Examples of social media impacts on next-day morning traffic: (a) time-series evaluation for a representative road segment. Each point is RMSE over a split; (b) Detailed predictions within a split. Each point represents error of one traffic prediction; (c) explanation of predictions made by $c > 1$ traffic descriptor on Jul 22, 2014; (d) explanation of predictions made by $c > 2$ traffic descriptor on Jul 22, 2014; (e) word cloud in the evening of Jul 21, 2014, and (d) word cloud in the late night of Jul 21, 2014.

*Social media data.* We show when social media features are necessary for next-day morning traffic prediction using a representative road segment (ID: 104N04503) as an example. The segment is located right after the merging lanes of William Penn Hwy and Penn Ave into I-376 W and is upstream to Squirrel Hill Tunnel, which is one notorious congestion bottleneck in Pittsburgh with extremely large day-to-day congestion variation. Note that this segment is located at the upstream edge of the traffic spillback area of Squirrel Hill Tunnel, indexed by TMC order = 1 in the cluster plots of I-376 W in Fig. 4. With tweet features approximating morning travel demand, this segment is an ideal location to test the effectiveness of social media data, since the limited supply only necessitates congestion while morning travel demand plays a crucial role in causing congestion on that segment.

As shown in Fig. 18, we first compare the prediction errors of models with and without tweet features by testing split in **(a)**. The model with tweet features performs much better than its variant for the test split in July. Thus, we zoom into the split in **(b)** and select 7/22/2014 (Tuesday) as a representative sample for examination. In the prediction made by thee road-level traffic descriptor, we visualize the contributions[7] of the top 5 features in **(c)** and **(d)**. Our model with tweets describes morning congestion with cluster-level 2 while the variant without tweets predicts cluster level 3 which misleads the prediction. Hence, we further examine the predictions made by $c > 1$ and $c > 2$ classifiers of the traffic descriptor with tweets in **(c)** and **(d)**. It is found that the percentages of neutral tweets in the evening and late night can trigger a prediction of level 2 morning congestion as in **(c)**, but are not high enough to predict level 3 congestion as in **(d)**. It indicates that some abnormal events which occurred in the evening and night of July 21 possibly reduced or put off morning travel demand on July 22. Finally, we confirm it by mining the topics of tweets posted in the two periods. The word cloud **(e)** of tweets posted in the evening, which contains "pnc park", "scoreboard", "pittsburgh pirates", "game" as frequent words, together with word cloud **(f)** of tweets posted in late night, which contains emotional words such as "love", "good", "shit", etc., and game-related words, indicate that there was a night game on July 21. In summary, we show that although our method does not improve morning traffic prediction for all road segments and days in our study, 78% of testing samples can benefit from the incorporation of social media by improving at least one of the four congestion measurement predictions.

*Traffic event information.* For all segments, only testing days with traffic incidents, which are defined by nonzero summation of incident features over all hours, are considered to assess the effectiveness of incident data reported before 5am for morning traffic prediction. As shown in Fig. 19, the performance degradation after removing incident features vary significantly by spatial locations of road segments. Traffic predictions of road segments adjacent to traffic bottlenecks (e.g., Ft. Pitt Tunnels, 40th St Bridge, Summer Hill, etc.) are generally more affected by the removal of traffic incident features. This is expected because traffic incidents occur frequently near traffic bottlenecks and are more likely to impact the traffic near the bottleneck and its upstream spillover. Interestingly, we also find in Fig. 19 that predictions of congestion status (ΔAccuracy) and congestion starting time (ΔCST) on segments upstream to these identified bottlenecks seem to be more affected by traffic incident features, while for predictions of CD and PTI, road segments upstream and downstream to traffic bottlenecks are equally affected. In fact, the results conform to the physical properties of network traffic flow because (1) congestion caused by traffic incidents propagates backward so all congestion measurements of segments upstream to traffic bottlenecks are affected by traffic incidents, and (2) traffic incidents can also block the outflow so segments downstream to traffic bottlenecks may be less congested than normal. Hence, predictions of CD and PTI for segments upstream to traffic bottlenecks can be affected by traffic incidents occurring at the bottleneck.

[7] The contribution of a feature to a prediction made by linear models are the product of feature value and feature weight.

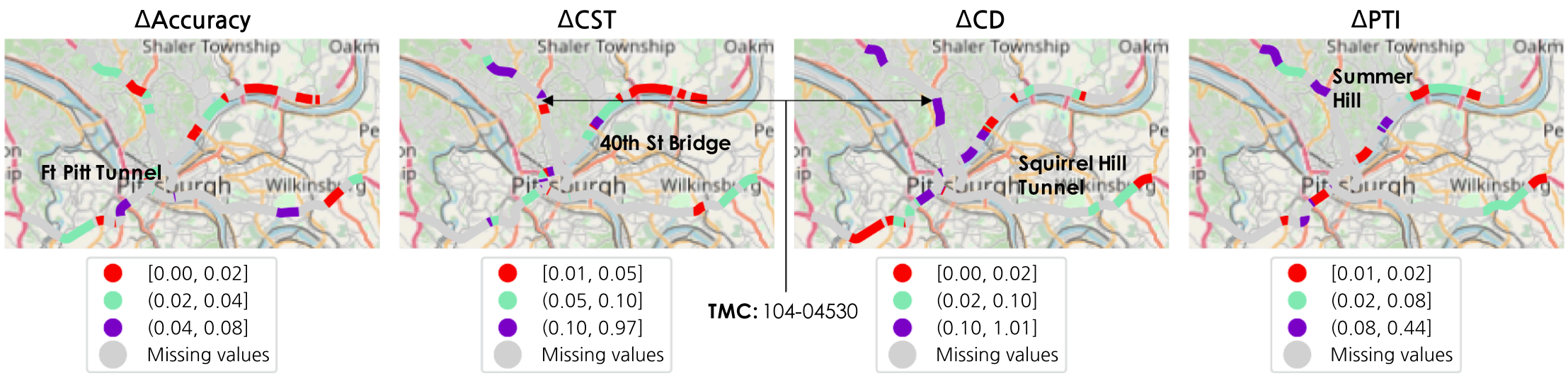

**Fig. 19.** Percentage performance degradation on non-recurrent incident days after removing traffic incident features.

We further examine a representative segment (ID: 104–04530) to support our finding. As shown in Fig. 19, the selected segment is close to Summer Hill where two highways I-279 SB and US-19 SB merge into Parkway North of Pittsburgh. The influences of traffic incident features on traffic prediction are crucial with $\Delta CST = 29.8\%$ and $\Delta CD = 64.4\%$. We fit the segment-level model and visualize the model weights of CST and CD predictors in Tables 5 and 6. For the CST predictor in Table 5, we show that traffic incidents occurring downstream at 7am (**p_ds_7**) bring forward CST while incidents downstream at 10am (**p_ds_10**) delay CST. Traffic incidents which partially closed the segment at 6am (**p_in_6**), or upstream at 4 and 9am (**p_us_4**, **p_us_9**) also delay the start of congestion. For the CD predictor in Table 6, a similar pattern is observed except that upstream incidents of all time periods (**p_us_Hour**) reduce the congestion duration. The fitted models support our proposed reasons (1) and (2) by explaining how traffic incident features impact morning traffic.

*Weather conditions.* In this study, we assume adverse weather conditions are comprised of snowing, raining, and wet pavement conditions. Hence, testing days with a nonzero summation of these features across all hours are considered to assess the effectiveness of weather data known before 5am for morning traffic prediction. Our results show that `tweet2traffic-no-weather` suffers from 3.5%±1.9% performance degradation in terms of prediction accuracy of CS, 27.6%±19.9% degradation in terms of RMSE of CST, 22.2%±13.0% for CD and 21.5%±13.3% for PTI on the same group of 10 road segments in Section 5.2.2. Compared with the results in Fig. 17, weather features are now crucial factors for predicting morning traffic, especially for CST, CD and PTI on days with adverse weather conditions. Note that we do not further analyze the contributions of weather data for different spatial locations because weather features are the same for all road segments. Also, we do not have enough segment-level attributes such as pavement types and slopes to explain the variances of precipitation impacts across the network.

### 5.3. *Effectiveness of clustered model structure*

We conduct experiments with `tweet2traffic-no-cluster` variant to show the effectiveness of the clustering in the proposed model structure. As shown in Fig. 20, after removing the clustered structure and feeding tweet features directly to segment-level models instead, most road segments suffer from performance degradation especially in terms of prediction errors of CST, CD and PTI. We also observe that the performance degradation is more significant for road segments in the middle of the road. For example, the RMSE of CST prediction for the segment in the middle of PA-28 SB increases by 80% after removing the clustered structure.

Our method which adds road traffic cluster probabilities into segment-level models resembles multi-level random intercept model (Raudenbush and Bryk, 2002), where the road congestion clusters can be considered as the random intercepts of upper groups (congestion clusters) and the features selected via $L_1$ penalty in segment-level models explain relationships in lower groups. As shown in Fig. 20, adding clustered structure is not always beneficial for segment models such as for segments on both ends of the road, which is reasonable as traffic at these locations does not follow overall congestion clusters closely. However, for segments in the middle of the road, model generalization performances have been largely improved because overall congestion clusters help segment models select critical features for explaining segment variances from high-dimensional features with a small sample size. The results also suggest that social media data seem to explain overall road-level congestion better than for specific locations.

### 5.4. *Sensitivity analysis: how far ahead morning traffic be well predicted with tweet information*

In previous sections, we show that most useful social media features are extracted from evening (EV) and late night (LN), which are readily available by midnight for the next-day traffic prediction. We thus conduct experiments with two `tweet2traffic-before-time` variants that use only features available before 3am or midnight. Specifically, for social media features, sleep-wake patterns and event indicators later than the cut-off time are removed. The same applies to weather features. However, traffic incident features remain the same because we assume planned traffic events are known beforehand. To prevent data leakage in modeling training, all selected features are re-normalized with the available data for each experiment.

As shown in Fig. 21, it is found that our method is generally insensitive to the increase of forecasting horizon for most road segments when tweet data are mined and used. This implies that we can even make the next-day morning traffic prediction by the midnight before, which is particularly appealing for proactive traffic management during morning commute. A larger performance drop is observed when increasing the horizon from "before 5am" to "before 3am", possibly because data between 3-5am captures people's waking-up patterns which can help explain morning traffic. However, the changes in performance with respect to forecasting horizons are relatively small, especially for CST, CD and PTI. For traffic management purposes, the optimal application of our method is to predict next-day morning traffic by midnight, which is expected to give commuters a sufficient time window to make trip decisions before/during morning commute.

**Table 5**
Segment-level model weights for CST prediction on 104–04530 ($L_1 = 0.6; R^2 = 0.543$). Details of features are in Table 3.

| Feature | Weight | Feature | Weight |
|---|---|---|---|
| BIAS | 38.394 | **p_ds_10** | −0.964 |
| **p_ds_7** | 0.518 | vis_5 | −0.491 |
| precip_hrly_5 | 0.355 | **p_us_4** | −0.474 |
| wx_phrase_5 | 0.127 | **p_us_9** | −0.197 |
| c_2 | 0.098 | | |
| precip_hrly_4 | 0.086 | | |
| pave_cond_4 | 0.052 | | |
| **p_in_6** | −2.246 | | |

**Table 6**
Segment-level model weights for CD prediction on 104–04530 ($L_1 = 0.6; R^2 = 0.460$). Details of features are in Table 3.

| Feature | Weight | Feature | Weight |
|---|---|---|---|
| BIAS | 8.286 | **p_ds_9** | 0.006 |
| c_2 | 2.028 | **p_us_5** | −0.001 |
| c_3 | 1.843 | **p_us_6** | −0.001 |
| precip_hrly_5 | 0.823 | **p_us_4** | −0.054 |
| **p_ds_8** | 0.728 | **p_us_9** | −0.242 |
| precip_hrly_2 | 0.128 | **p_ds_5** | −0.391 |
| wx_phrase_1 | 0.101 | vis_3 | −0.496 |
| precip_hrly_4 | 0.018 | vis_5 | −0.55 |

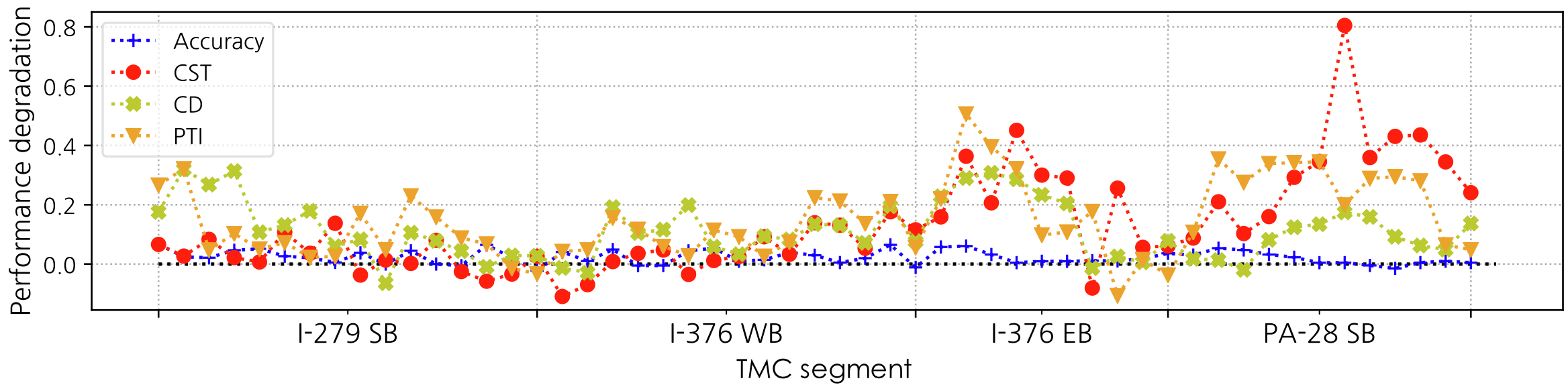

**Fig. 20.** Percentage performance degradation after removing clustered model structure.

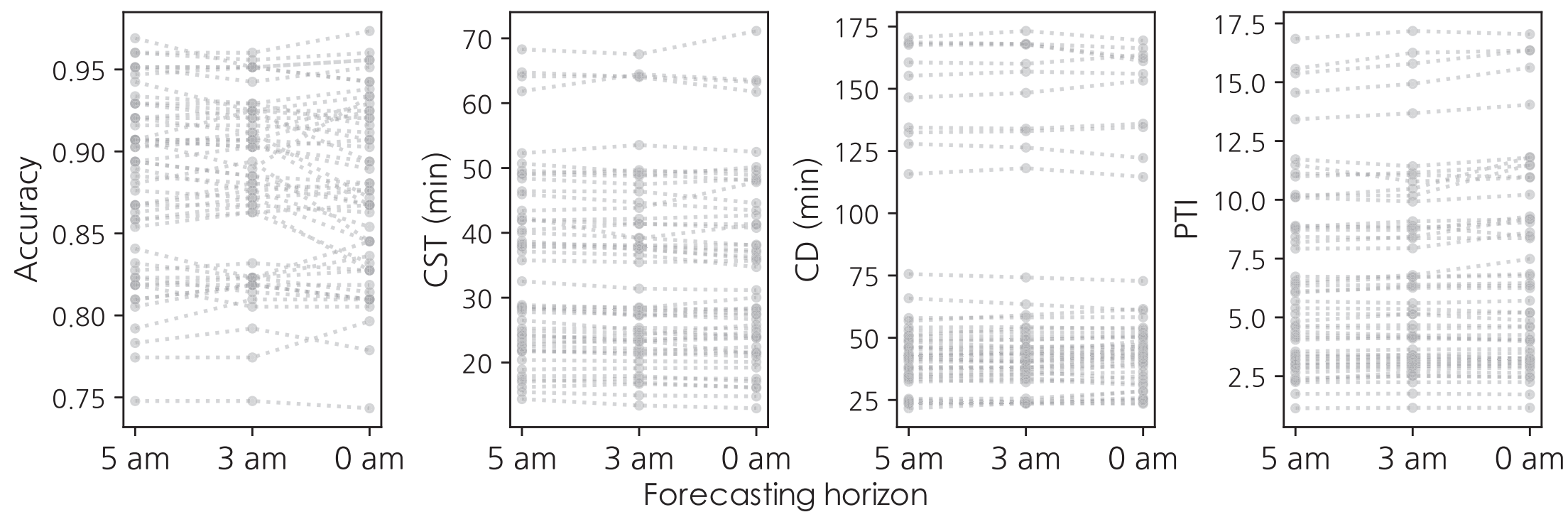

**Fig. 21.** Next-day morning traffic prediction errors by road segment under three forecasting horizons. Each line represents a road segment and each point is model prediction error with a specified forecasting horizon on x-axis.

## 6. Conclusions and future work

This paper demonstrates the possibility of using social media data that are publicly available online as a crowdsourcing tool to predict next-day morning traffic. A comprehensive framework, namely `tweet2traffic`, has been designed for regional networks which first characterizes morning road congestion, processes and extracts traffic information from social data, then examines the relationship between social media and morning congestion patterns, and ultimately uses the relationship to predict morning traffic using data by midnight or up to 5am. Specifically, we extract three types of information from tweets for explaining next-day morning traffic, which include people's sleep-wake status, local events, and planned traffic incidents for next day. Considering the sparsity of geocoded tweets, we propose a novel social media data augmentation method which first filters local residents from noisy users such as tourists, infers their home locations, and augments the dataset with user timelines such as non-geocoded tweets, retweets, and favorites by assuming residents stay at home during the night. A sentiment analysis model with a neural language model as backend is constructed to detect abnormal local events and incident records are extracted from traffic-related tweets. To remove compounding errors, we propose to describe morning traffic with a quadruple of variables including binary congestion status, congestion starting time, congestion duration, and planning time index. A traffic prediction pipeline which consists of three building blocks: (1) road-level traffic descriptor, (2) segment-level traffic classifier and (3) segment-level traffic regressor, is developed to encode the supply-side variables (traffic incidents, weather, etc.) and demand-side variables (tweets, temporal variables, etc.) through a clustered learning structure which makes use of ordered spatio-temporal congestion patterns of segments along a road.

Through experiments in the Pittsburgh region, we show that our approach outperforms existing baseline methods and can learn meaningful travel demand characteristics from tweeting profiles and activities. Generally, we discover that the earlier people go to sleep, the more congested the road will be in the next morning. Tweeting activities in the night before and early morning are statistically associated with morning congestion. In addition, we find that the occurrence of big events in the evening, represented by higher or lower tweet sentiment than normal, often implies lower travel demand in the next morning than normal days. We illustrate through ablation studies that social media data, traffic events, and weather conditions are important data sources for next-day morning traffic prediction and their roles in prediction vary by space and time. We find that tweet features can effectively improve next-day morning traffic prediction for some road segments, particularly upstream of the pillback areas of a typical bottleneck and with large day-to-day congestion variation, where congestion patterns can be very sensitive to travel demand characteristics. In addition, the impacts of social media data are more significant when special events, such as sports games, occur during the night, which possibly reduces or puts off morning travel demand. In addition, information of traffic incidents and weather conditions improves predictions on days of non-recurrent traffic with roadworks or adverse weather conditions. Traffic incidents downstream to road segments shift congestion starting time while upstream incidents reduce congestion duration. Adverse weather conditions often increase congestion scales. Finally, our approach proves to be insensitive to the increase in forecasting lag times, and multi-source data up to midnight can be sufficient in providing good prediction of next day morning traffic. This is particularly appealing for traffic management, as traffic information and management strategies can be provided to commuters proactively before the morning starts.

Our experiments suggest that the multi-source data features, including tweets, incidents, weather and temporal variables, are only effective for improving morning traffic predictions on some locations in the network, but not for all. This is completely anticipated as some road segments are much more likely to be influenced or implied by weather, population activities and incidents than others. The data driven approach precisely helps learn the "signals" of morning traffic congestion as much as it can from social media data, which is effective and powerful on a portion of roads. This `tweet2traffic` model may not work for other roads, simply because social media are not actively used in areas that are relevant to those roads.

Last but not least, it is important to note the limitations of the current study. First, the main task of this study is to improve morning traffic prediction with social media data and travel demand characteristics are interpreted from the linear predictive models. Hence, no causal relationship can be claimed by this paper. Second, the loss of Twitter users with geo-coded tweets has been observed over the past few years. It's therefore worth investigating how our method can still work using the active tweet users that can change considerably over time. Finally, some road segments in our study are shown to have high prediction errors even using social media. In a nutshell, integrating social media data into traffic prediction opens up a wide range of opportunities for transportation research to mine cross-domain data sources for predicting short-term traffic. In our future work, we plan to add new types of cross-domain data collected from buildings (Chen et al., 2020), energy systems (Zhang and Qian, 2018), water/sewer systems, city light to predict long-term traffic, because all those data can strongly imply population activities relevant to travels. Examining the impacts of the changing tweet users and topics to traffic prediction over time is also an interesting direction.

## CRediT authorship contribution statement

**W. Yao:** Data curation, Formal analysis, Investigation, Methodology, Software, Supervision, Validation, Visualization, Writing - original draft, Writing - review & editing. **S. Qian:** Conceptualization , Formal analysis, Funding acquisition, Investigation, Methodology, Project administration, Resources, Supervision, Validation, Visualization, Writing - original draft, Writing - review & editing.

## Acknowledgements

This project is funded by National Science Foundation Award CMMI-1751448, and in part by Carnegie Mellon University's Traffic21 Institute and Mobility21, a National USDOT University Transportation Center for mobility sponsored by the U.S. Department of Transportation.